\documentclass[manuscript]{acmart}
\AtBeginDocument{%
  \providecommand\BibTeX{{%
    \normalfont B\kern-0.5em{\scshape i\kern-0.25em b}\kern-0.8em\TeX}}}

\begin{document}

\title{Assessing the relationship between subjective trust, confidence measurements, and mouse trajectory characteristics in an online task}
%


\author{Martin Dechant}
\email{martin.dechant@zeiss.com}
\orcid{0000-0001-9073-8727}
\affiliation{%
  \institution{Carl Zeiss Vision International GmbH}
  \streetaddress{Turnstrasse 27}
  \city{Aalen}
  \country{Germany}
  \postcode{73430}
}

\author{Susanne Poeller}
\email{susanne.poeller@usask.ca}
\orcid{0000-0002-6930-5318}
\affiliation{%
  \institution{Department of Computer Science, University of Saskatchewan}
  \streetaddress{P.O. Box 1212}
  \city{Saskatoon}
  \state{Saskatchewan}
  \country{Canada}
  \postcode{S7N 5C9}
}
\author{Benedikt Hosp}
\email{Benedikt.Hosp@uni-tuebingen.de}
\orcid{0000-0001-8259-5463}
\affiliation{%
  \institution{ZEISS Vision Science Lab, Institute for Ophthalmic Research, University Tübingen }
  \streetaddress{Elfriede-Aulhorn-Straße 7}
  \city{Tübingen}
  \country{Germany}
  \postcode{72076}
}

\author{Olga Lukashova-Sanz}
\email{olga.lukashova@zeiss.com}
\orcid{0000-0002-4053-3894}
\affiliation{%
  \institution{Carl Zeiss Vision International GmbH}
  \streetaddress{Turnstrasse 27}
  \city{Aalen}
  \country{Germany}
  \postcode{73430}
}

\author{Alexandra Sipatchin}
\email{Alexandra.Sipatchin@uni-tuebingen.de}
\orcid{0000-0002-2768-1040}
\affiliation{%
  \institution{ZEISS Vision Science Lab, Institute for Ophthalmic Research, University Tübingen }
  \streetaddress{Elfriede-Aulhorn-Straße 7}
  \city{Tübingen}
  \country{Germany}
  \postcode{72076}
}

\author{Siegfried Wahl}
\email{siegfried.wahl@zeiss.com}
\orcid{0000-0003-3437-6711}
\affiliation{%
  \institution{Carl Zeiss Vision International GmbH}
  \streetaddress{Turnstrasse 27}
  \city{Aalen}
  \country{Germany}
  \postcode{73430}
}

\renewcommand{\shortauthors}{Dechant, et al.}

\begin{abstract}
Trust is essential for our interactions with others but also with artificial intelligence (AI) based systems. To understand whether a user trusts an AI, researchers need reliable measurement tools. However, currently discussed markers mostly rely on expensive and invasive sensors, like electroencephalograms, which may cause discomfort. The analysis of mouse trajectory has been suggested as a convenient tool for trust assessment. However, the relationship between trust, confidence and mouse trajectory is not yet fully understood. To provide more insights into this relationship, we asked participants (n = 146) to rate whether several tweets were offensive while an AI suggested its assessment. Our results reveal which aspects of the mouse trajectory are affected by the user’s subjective trust and confidence ratings; yet they indicate that these measures might not explain sufficiently the variance to be used on their own. This work examines a potential low-cost trust assessment in AI systems.
\end{abstract}



\maketitle

\section{Introduction}





Trust plays a crucial role not only during the interaction with other individuals \cite{Zhu2019RivalsHiding} or organizations \cite{Arrow1974TheOrganization} but also in the success of a device or interface \cite{Sollner2016WhyUsers}: While information systems aim to make the user's life easier, these systems may also become increasingly hard to understand due to the complexity of the underlying technology. As Narbona \cite{Narbona_FragileTenetsOfTrust} et al. emphasize Trust is a fragile construct and once lost, it becomes quite challenging to restore trust. As a result of the lack of trust in a system, users may feel frustrated or even intimidated by the complexity \cite{Coskun2005ImpactsSystems,vanderWaa2018ContrastiveConsequences} and start to lose their trust in the system \cite{Burton2020AMaking}. In extreme cases, users may stop using the system completely since they mistrust it and would rather rely on their own intuition, as prior works shows \cite{Tolmeijer2022CapableMaking,Wu2022SafetyNavigation}: mistrust may have lead severe accidents by ignoring the AI's warning \cite{Bliss2003AlarmDriving} or making dangerous decisions \cite{Xiao1999AnTasks}.
In recent years, developers started to improve various devices and applications by embedding artificial intelligence. On the one hand, AI-based systems have been improving the quality of many lives in various situations like online shopping \cite{Bawack2022ArtificialReview}, media consumption \cite{Lachman2021ApplicationsEntertainment} and entertainment industry \cite{Chan-Olmsted2019AIndustry}, or even within the realm of health care and well-being \cite{Yu2018ArtificialHealthcare}. Not only are AI-based systems affecting our private lives, but they also rapidly change our work environment by introducing smart agents and systems to increase performance by assisting users in their daily routines \cite{Damioli2021TheProductivity}. 
However, AI also needs to earn the user's trust \cite{Glikson2020HumanResearch,Bockle2021CanInterfaces} similar to other technologies, when they entered and changed our the daily life, such as the smartphone revolution. While artificial intelligence offers many benefits, there are still many concerns and mistrust in AI which may even cause in severe cases anxiety \cite{Johnson2017AIAnxiety}. This problem is intensified by a misunderstanding of the capabilities and a lack of understanding of the underlying technologies. As a result of this lack of knowledge, users may understand artificial intelligence as a "black box", implying an underlying scepticism towards the technology \cite{Bockle2021CanInterfaces}. Recently, AI systems were heavily criticized for being, for many users, too complex and not understandable for users \cite{BarredoArrieta2020ExplainableAI}.To counteract this problem researchers and designers of AI systems began to adapt the principles of explainable AI  \cite{BarredoArrieta2020ExplainableAI,Gunning2019XAIExplainableIntelligence}: the idea of explainable AI proposes to focus not only the system's performance but also the comprehensibility of the system during the development processes. This emphasis on explainable AI procedures may help users build trust in the system and help developers effectively manage the user's expectations at the same time \citep{Kaur2023TrustworthyReview}. 
While this approach may play an important role for the success of AI, one question challenges AI developers: \textit{how can we measure the user's trust in the AI?}

As prior work outlines, trust should be seen as an attitude, meaning that even if we experience trust, we may not show easy to measure phyiological reactions,  and not as an emotional state \cite{Dewey1895TheEmotion.}, like for example anxiety\cite{Johnson2017AIAnxiety,Dechant2021TheAnxiety,Rinck2010AttentionalSpiders}, which may challenge how to empirically measure trust\cite{Vereschak2021HowMethodologies} . 
In the context of trust measurements, prior work suggests a broad variety of standardized questionnaires and ratings as a way to assess the construct \cite{Vereschak2021HowMethodologies,Kohn2021MeasurementGuide,Leichtenstern2011PhysiologicalSituations,Gupta2020MeasuringReality}.Furthermore, there are several optimized questionnaires, which either focus on trust in other individuals \cite{Wrightsman_1991} or technology in general \cite{Merritt2013ISystem} in order to capture the context-sensitivity of trust. However, there are several challenges with these approaches: On one side there is an ongoing discussion whether simple subjective rating-based approaches may actually be enough to capture the user's trust \cite{Kohn2021MeasurementGuide,Chita-Tegmark2021CanMeasure}. Furthermore, delivering questionnaires may disturb the main task and distract users, which makes it hard to measure trust during an interaction \cite{Lu2019EyeReliability}. To overcome these problems, researchers introduced the principles of evidence-based assessment, which combines subjective measurements and behavioural data-assessment\cite{Hunsley2007Evidence-BasedAssessment,Mash2005Evidence-basedChallenges}.
To provide evidence for the experience of trust, researchers began to look into sensory and behavioural data for the development of behavioural and digital biomarkers \cite{Strimbu2010,international2001biomarkers}. Digital biomarkers, described as objective and quantifiable data collected via digital devices, may strengthen the assessment of trust by increasing the temporal and spatial resolution of captured behaviour during the assessment \cite{Mandryk2019}. Therefore, researchers have begun to collect physiological measurements such as heart rate \cite{Merrill2017TrustInteractions},galvanic skin conductance \cite{Morris2017ElectrodermalUse}, and brain activities using Electroencephalography (EEG) \cite{Akash2018AGSR}, and gaze data \cite{Barbato2020TheOrienting}. While these investigations provide promising results, they require expensive hardware or complicated procedures but also the presence of complex hardware may intimidate users \cite{Fessl2011MotivationReflection,Ahlan2014UserModel,vanderWaa2018ContrastiveConsequences} and lead to a loss of trust due to increased complexity.
Another proposed way of assessment is the analysis of trust-related behaviour measurements, like response time \cite{Hu2016Real-TimeFoundation.}, compliance (the number of times participants followed the system's recommendations)\cite{Brzowski2019TrustReview}, reliance (the number of times participants asked for support from the system) \cite{Sutherland2015TheAutomation,Vereschak2021HowMethodologies}, and the switch ratio (how often participants decided to give up their own opinion and follow the AI's recommendation). 
One not yet fully explored approach is the analysis of the mouse trajectory \cite{Vereschak2021HowMethodologies}. Adjunct research shows, that mouse trajectory is affected by hesitation \cite{Zushi2015AnalysisProblems,Arroyo2006UsabilityTracks}, cognitive load \cite{Rheem2018UseLoad} or even anxiety \cite{Yamauchi2013MouseForest}. However, a clear analysis of the relationship between trust and adjunct phenomena such as confidence in their decision on the mouse trajectory is still missing. By providing a better understanding of this relationship, future researchers and developers may be able to use mouse trajectory as a cheap and easily accessible tool for assessing trustful behaviour. 
To inform our understanding of the relationship between trust, confidence and expression through mouse trajectory, we asked participants (\textit{n}=146) to engage in a series of collaboration tasks with two simulated AI agents in a rating task, which was adapted from prior work \cite{Papenmeier2022ItsAI} for the assessment of the mouse trajectory. The AI either had an accuracy of 90\% or 60\%.Participants were instructed to rate whether a presented short message was offensive or could be published on a social media platform \cite{Jay2008TheSwearing,Papenmeier2022ItsAI}. In each trial, the AI showed a recommendation which was either "right" or "wrong", depending on the indicated accuracy. An high accuracy means in this context that the AI would suggest more often the correct answer. After each trial, we asked participants to rate their experienced subjective trust in the AI system and their confidence in their decision. 

Our findings show that individuals who indicated an elevated level of trust in the self-report-measure  spent less time on the task and had a shorter mouse path, and the average and maximum mouse speeds were higher. Similar results were also found for self-reported subjective measurements of confidence in their answer. While these results are promising, the resulting models were only able to explain a small portion of the data's variance in our sample, which means that the subjective measurements of trust affected the mouse trajectory characteristics little. However, the subjective measurement of confidence had similar influence on the mouse trajectory. Both aspects emphasize that using additional measures, subjective and objective ones, is necessary to better understand whether the user trusts the system or was confident in their own decision. By showing strengths and weaknesses of this assessment tool, we help researchers and designers to refine their evaluation procedures to test whether a user trusts the AI.

\section{Related Work}
We will first present key characteristics of trust, then provide an overview of the distinction of trust to adjunct phenomena and finally discuss currently used ways for assessing trust.

\subsection{Definition of Trust}
 As prior work outlines \cite{Vereschak2021HowMethodologies,Ueno2022TrustMethods} trust plays an essential role in the acceptance of technology, researchers have investigated which aspects alter the manifestation of the user's trust during the interaction with the AI. Acceptance of technology is not only affected by the general performance and reliance on the system \cite{Zhang2020EffectMaking} but also the perception of safety \cite{Wu2022SafetyNavigation,OCass2003WebBehaviour}, ease of use \cite{Davis1989PerceivedTechnology} and compatibility with the personal environment \cite{Hernandez-Ortega2011TheConsequences}. Further, the user's cultural background may affect the manifestation of trust \cite{Doney1998UnderstandingTrust}. Due to the rising popularity of artificial intelligence enhanced systems, researchers also started to investigate in particular the relationship between the user's trust and characteristics of artificial intelligence as well as ways to enhance the human-AI interaction by strengthen the user's trust\cite{Kaur2023TrustworthyReview}. 
 
 Similar to that question other research disciplines have investigated as well the concept of trust in their domain like in professional environments \cite{Zaheer1998DoesPerformance}, social relationships \cite{Sutcliffe2015ModellingRelationships} but also the importance of trust in media \cite{Schranz2018MediaUse} and government \cite{Chanlet_Consequences_Trust_Government}. Therefore multiple definitions of trust emerged \cite{Papenmeier2022ItsAI,Rousseau1998NotTrust}, each with their own perspective on trust. The recent systematic review of Vereschak et al., which focuses on the context of Human-Computer Interaction \cite{Vereschak2021HowMethodologies}, defines three major reoccurring components of trust: vulnerability, positive expectations, and the fact that trust should be seen as an attitude. Vulnerability and positive experience are described as basic conditions for trust\cite{Rousseau1998NotTrust}, and the idea that trust is an attitude further defines ways of investigating and measuring it.

The first key component of trust is the idea of vulnerability \cite{Lee2004TrustReliance}. It describes the experience of situations which involve uncertainty of the outcomes of a decision, for example, the possibility of a negative outcome or undesirable consequence \cite{Hosmer1995Trust:Ethics}. However, it's important to differentiate between the two types of uncertainty \cite{Knight1921RiskProfit}: we can estimate the outcome of a decision in some situations(e.g. crossing a road). This is sometimes described as risk versus ambiguity (e.g. gambling at a casino). However, without experiencing any vulnerability there is no need for the user to trust anyone else in the situation \cite{Lee2004TrustReliance,Rousseau1998NotTrust,Vereschak2021HowMethodologies}.

The role of expectations in human-AI interaction plays a crucial role in the experience of trust. In particular, trust may only manifest when an individual has the expectation that another entity will be able to perform a benevolent action. This includes that the individual believes that by trusting the other entity, any negative consequences are at least partly avoided \cite{vanMaanen2011EffectsAutonomy,Rajaonah2006TrustControl}. 

However, it is also important to emphasize that trust is defined as an attitude \cite{Yu2019DoDecision,Papenmeier2022ItsAI} meaning that the experience of trust does not systematically translate into behaviour and could be overwritten by other inputs. For example: a user may trust a suggested route of a navigation system but still not follow the recommended route due to other information, e.g. trusting their own knowledge over shortcuts.

Moreover, trust itself can be distinguished by two different concepts: on the one hand, trust can be seen as dispositional trust. This describes a personality trait like the tendency to trust or mistrust others \cite{Merritt2008NotInteractions}. On the other hand, we have the concept of situational trust, which is based on the impression and interaction with another entity, like a colleague or artificial intelligence.

\subsection{The Relationship between Trust and Confidence}
All three previously introduced components need to be present in order for trust to emerge. However, there are overlapping concepts with trust \cite{Vereschak2021HowMethodologies}, mostly when one of the components is missing. As mentioned above, trust manifests in the idea of vulnerability in a task. Nonetheless, researchers argue that the lack of vulnerability in a situation may lead to the experience of confidence \cite{,Evans2009TheTrust,Luhmann1988Trust:Relations}. Confidence describes the feeling of an individual's security regarding \textit{their own decision} \cite{Evans2009TheTrust,Luhmann1988Trust:Relations,See2011TheAccuracy}. For example: A patient decides to follow a treatment without considering any alternatives and thinks that they
will only be better off which may suggest that the user experience confidence in their decision \cite{Vereschak2021HowMethodologies}. Some researchers even argue, that trust should be seen as a sub-category of confident behaviour,, and therefore define trust as placing confidence in another entity's decision while being vulnerable to the consequence of this decision \cite{Ryan2020InReliability}. This emphasizes the blurry line between the experience of trust and being confident in the own decision. However, both concepts may lead to different interactions, where trust leads to a cooperation with the system confidence in the own decision may lead to a focus on the own intuition and ignoring the suggestions of the AI.

\subsection{Measurements of Trust}
As previously discussed, the attitude characteristic of trust makes a reliable assessment of the user's trust in a system quite challenging \cite{Papenmeier2022ItsAI,Yi2020IdentificationEmotions}. To overcome this problem, researchers apply different approaches and technologies ranging from the use of standardized questionnaires to the measurement of behaviour linked to potential trustful interaction \cite{Vereschak2021HowMethodologies}. However, there is still an ongoing debate about whether such approaches are valid \cite{Papenmeier2022ItsAI,Vereschak2021HowMethodologies}. Different contexts may require different aspects of trust to be measured. As previously discussed: trust can either focus on a device in a specific moment or the general trust in a type of technology \cite{Wrightsman_1991,Merritt2013ISystem}.

\subsubsection{Questionnaire-Based and Subjective Assessment of Trust}
Standardized questionnaires and ratings are one commonly used way to assess the subjective view of the users and allow them to rate whether they experience trust. Most of these questionnaires depend on the context and the underlying research question. For example: \citet{Merritt2013ISystem} focuses on the overall trust in technology, while other questionnaires, like that of \citet{Wrightsman_1991}, focus on whether an individual generally trusts other people \cite{Wrightsman_1991}. Another way to assess trust is a simple rating measurement, where users simply indicate how much they trusted a system \cite{Yin2019UnderstandingModels}. However, some researchers argue that these questionnaires and rating-based measurements of trust are not able to capture the complexity of trust \cite{Loo2002AScales}. Other researchers argue that trust should not only be measured on its own and recommend combining additional measurements to distinguish between adjunct phenomena \cite{Vereschak2021HowMethodologies}.  

\subsubsection{Behavioural Markers and Physiological Data for Assessing Trust}
Besides the use of questionnaires, researchers have started to harness additional data sources to find ways to assess trust in different contexts, including the human-AI interaction \cite{HarrisonMcKnight2001TrustTime}. However, these may be among the most affected by the characteristics of trust, meaning that trust may not be directly expressed through measurable behaviour or signals \cite{Castelfranchi2010_Socio}. In comparison: emotional states such as anxiety may cause distinct physical reactions \cite{Spence2016} such as trembling \cite{Connor2000PsychometricScale} or increased pulse \cite{Cheng2022HeartMetaanalysis}. Therefore, prior work suggests \cite{HarrisonMcKnight2001TrustTime,Castelfranchi2010_Socio,Vereschak2021HowMethodologies} describing these measurements as trust-related behavioural measurements to account for the fact that these measurements may not fully represent trust. 

Researchers in related contexts, such as mental health assessment, coined the term "digital biomarkers", which describe behavioural data which may correlate with a mental state or illness and can be used in combination with subjective measurements as a way to assess the user's mental health \cite{Strimbu2010,Mandryk2019}. Within the context of trust-related behaviour, prior research describes similar potential measurements, such as the decision time, i.e. how fast a user follows an AI's recommendation \cite{Hu2016Real-TimeFoundation.} or how often a user followed such recommendations, which is referred to as compliance \cite{Brzowski2019TrustReview}. Additionally, researchers investigated whether the reliance \cite{Sutherland2015TheAutomation,Vereschak2021HowMethodologies}, namely the number of user requests for AI guidance, may be useful for this assessment, as well as the switch ratio, time, and the number of trials in which the user changed their mind due to the AI's recommendations \cite{Vereschak2021HowMethodologies}. 
Besides these basic aspects, researchers have begun to look into physiological measurements, like the usage of heart rate variability \cite{Merrill2017TrustInteractions}, galvanic skin responses \cite{Morris2017ElectrodermalUse} or even harnessing Electroencephalography (EEG) \cite{Akash2018AGSR} to gain insights into the manifestation of trust during an interaction. 

\subsection{Context of this study}
Understanding whether a user trusts artificial intelligence is crucial for an efficient and productive interaction. However, this measurement can be quite challenging due to the characteristics of trust. Therefore, we address the need for solid ways to measure trust by analyzing potential behavioural metrics and their relationship to trust and trustful behaviour. While these metrics sound promising, some uncertainty remains. First of all, while prior work suggests the mouse trajectory as a potential trust-related behaviour, evidence is missing. Second, most of these studies investigated only trust, yet related concepts, such as confidence, are still under-explored. Therefore the relationships between trust, confidence, and potential behavioural data, are unclear. Thus, more research is required to understand this relationship. To address this gap we conducted an online experiment in order to provide insights into the relationship between the mouse trajectory, trustful behaviour, and the experience of confidence.

\section{Methods}
To understand whether the experience of trust, as well as confidence in a human artificial intelligence interaction context, affects the user's behaviour in a characteristically way, we conducted an online study in which participants had to interact with a simulated AI-enhanced agent with two different levels of accuracy. As prior work suggests, we adapted a trust assessment task \cite{Papenmeier2022ItsAI} to assess characteristics of the mouse trajectory.
\subsection{Apparatus and Conditions}
Since prior work \cite{Yin2019UnderstandingModels} shows that the presented accuracy of an AI may affect the user's trust, we investigated whether this also affects the manifestation of trustful behaviour. In this experiment we had two conditions: Low-accurate AI(60\%) and High-accurate AI(90\%).The percentage of the accuracy represents how many correct answers per condition were presented. For example, an accuracy of 90\% means that the AI would present a correct answer in 18 out of 20 trials. Since we used a within subject design we randomly assigned participants to one of two experimental orders of the two AI conditions at the beginning of the experiment. 
\subsubsection{The Task}
\begin{figure}[t]
\includegraphics[width=1\textwidth]{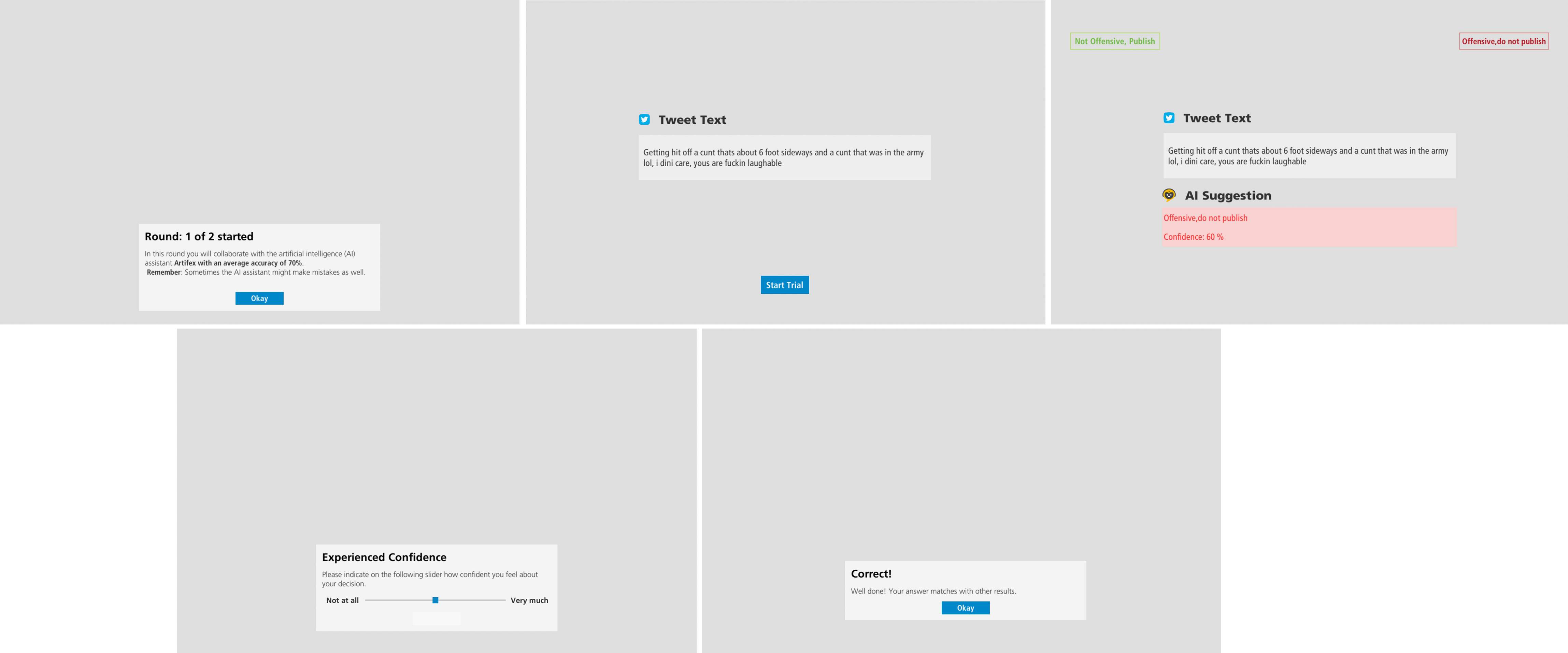}
\centering
\caption{The screens of the trail, from left to right: 1) The level of confidence of the AI is presented. 2)The tweet is presented; Once participants read it they can continue by clicking the button below 3)The AI suggestion is presented below in red or green below the tweet. The users can make their decisions on the top buttons  4)The user rates their experienced confidence and trust after selecting an answer 5) The answer is revealed whether the tweet was \textit{actually} offensive}
\Description{1) We see a grey screen with a white text box with lack text and a blue button describing the accuracy of the AI

2) A white text box in the center of the grey screen shows the tweet with a small twitter bird icon. Beneath the text box is a blue button to start the experiment. 

3) The experiment starts. The AI suggestion appears below the tweet. It is either in Red or Green depending of users should reject or accept this tweet. Two buttons, also red or green, appear on the top corners of the screen.

4) The tweet and the AI suggestion disappeared. Only a white box with a slider to assess the experienced trust appeared.

5) a white box with a slider to assess the experienced confidence appeared.}
\label{fig_Methods}
\end{figure}

In the task, which was adapted from prior work \cite{Papenmeier2022ItsAI,Yi2020IdentificationEmotions}, participants were asked to decide whether a presented problematic tweet was okay to publish and follows community standards on the platform or had to be removed. To analyze whether the user's mouse behaviour reveals insights about the experience of trust in an AI, we converted this task into a mouse trajectory assessment \cite{Freeman2018DoingHand,Kieslich2017Mousetrap:Package}, resulting in the following procedure: first, we showed participants the tweet in the centre of the screen as well as a button to continue the experiment on the lower part of the screen. Once they finished reading the tweet they had to press the button to start the trial. After starting the trial, the suggestion of artificial intelligence was presented below the tweet text. Depending on the condition, the AI suggestions were correct or suggested the wrong answer. Additionally, we showed an estimated accuracy value of the AI based on the condition. This information was added to maintain the illusion of a real artificial intelligence. Similar to the original task proposed by prior work \cite{Papenmeier2022ItsAI} the suggestion of the AI was either highlighted green (=\textit{"Not offensive, do publish"}) or red (=\textit{"Offensive,do not publish"}) as well as the two answer buttons \textit{"Offensive, do not publish"} and \textit{"Not offensive, do publish"} on the top left and right side of the screen, respectively. Similar to the AI suggestion, the buttons were highlighted in colour based on their text. The position of these buttons was flipped at 50 \% of the trials per condition. The flipping of the answer was used as an attention certification, where users who always selected one side over the whole experiment suggest that they may not pay attention to the experiment. We instructed players to move the mouse to the button and click to make their decision. There was no way to adjust their answer after pressing the button. After indicating their decision, participants were asked to rate their experienced trust in the AI's recommendation as well as their experienced confidence (see Measurements) in their decision. After that, we revealed whether the submitted answer was correct, namely matched the suggestions of other raters. In total, participants rated 20 tweets per condition. we showed the same tweets in each condition in randomized order to avoid any potential bias induced by the tweets. Furthermore the suggestions of the AI were randomly assigned based on the underlying accuracy of the current condition so that in one round the AI could suggest that the tweet was offensive while in the second round it suggested the opposite.  \autoref{fig_Methods} summarizes summarizes the task. 

The tweet texts were selected from a database of rated tweets, similar to prior work \cite{Papenmeier2022ItsAI,Jay2008TheSwearing}. We selected 20 tweets, which were assessed by multiple raters but not clearly identified as offensive. To prevent any bias induced by the content of the tweets, we adjusted the content. For example, we replaced gender-specific swearing with more neutral-themed ones. The task was implemented using the Unity Game Engine \cite{UnityTechnologies2020} and the Framework "Bride of Frankenstein" \cite{Johanson2020}.

\subsection{Participants and Procedure}
For this experiment, we recruited 182 Participants on the platform Prolific \cite{Prolific2014Prolific} who indicated that they are currently living in the United Kingdom and identified themselves as English native speakers in order to avoid any language-based misunderstandings of the presented content. After providing consent, participants were introduced to the controls and the study in the following way: 

\textit{"In this task you will see several controversial tweets which might be offensive or of bad taste. You have to decide whether these tweets are safe to publish or not. To make the decision easier for you, an artificial intelligence (AI) will assist you during the task and suggest whether a tweet is offensive or not. Each model is trained on real twitter data to give you the best recommendations as possible. However, even the best model may make mistakes and it is your decision whether you trust the AI's recommendations or not. In the end we will compare your answer with the suggestion of prior raters  and check whether your answer matches with the thoughts of the community. Your performance will help us reduce the hate speech online and create a more inclusive digital environment."} The order of conditions was randomized between participants and all participants had to perform both conditions.

After that, participants performed two exercise trials, in which two random tweets were shown and the AI always suggested the correct answer to illustrate the task procedure. Thereafter, the task started. We introduced the AI in a text message as well as the overall accuracy of the AI model. We repeated that for both levels of accuracy. After finishing both rounds of the task we asked participants to respond to several questionnaires about their general experience with technology, their general confidence, general trust, and trust in technology, as well as demographic information. Finally, we debriefed participants and offered crisis hotlines, and support resources in case participants experienced any negative feelings due to the exposure to potentially offensive language in the tweets.

\subsection{Measurements}
To better understand the relationship between trust in an AI and the potential manifestation through behavioural adaptations, we followed the suggestions of evidence-based assessment procedures \cite{Hunsley2007Evidence-BasedAssessment} and included subjective measurements as well as the potential behavioural markers.

\subsubsection{Rating and Questionnaire-based Measurement}
We used several ratings to assess the situational trust of the participants after each trial as well as the participant's dispositional trust and trust in technology: 
\begin{itemize}
\item \textit{Dispositional Trust Level:} We used the  \citet{Wrightsman_1991} questionnaire in order to measure the participant's dispositional trust in others. It asks participants whether other individuals are honest and reliable and whether they consider them moral. The scale consists of 14 statements, seven positive and seven negative. Participants indicate their rating on a 7-point scale, ranging from -3 to 3, how much they agree with the shown statement. All answers are summed up to a score, ranging from -42 up to 42. This score indicates how much an individual may trust another person.\par
\item \textit{Trust in Technology:} Besides the assessment of trust in \textit{others}, we also assessed the user's level of trust in machines. We used the propensity to trust scale by \citet{Merritt2013ISystem}. This scale consists of six statements. Participants indicated their agreement with each statement on a five-point Likert scale, ranging from "strongly Agree" to "strongly Disagree".\par
\item \textit{Personal Self-esteem:} Since we were also curious about the role of confidence in their own decision, we used the \citet{Rosenberg1965RosenbergScale} self-esteem scale to assess the confidence of the participants in their decision. The scale consists of 10 statements of both positive and negative feelings about oneself. Participants rated how much they agreed with these statements on a 4-point Likert scale, ranging from strongly disagree (1) to strongly agree (4). The sum of all items indicates self-esteem.\par
\item \textit{Experienced Trust of the AI}: After confirming the answer to the system, we asked participants to rate their experienced situational in the assistant using a Likert Scale from 0.0, not at all, to 1.0, very much in 100 steps.\par 
\item \textit{Confidence in their Decision}: We further asked participants to rate how confident they feel about their decision using a Likert Scale ranging from 0.0, not at all, to 1.0, very much in 100 steps.\par
\item \textit{Demographic Information:} We asked the participants to indicate age and gender as well as to summarize the experiment to check for (lack of) attention paid to the study. \par
\end{itemize}

\subsubsection{Behavioural Measurements} 
As suggested by prior work, we investigated several aspects of the mouse trajectory and adjunct data:
\begin{itemize}
\item \textit{Spent Time in Trial}: We measured time spent on the task, starting at the moment participants pressed the "start trial button" and stopping once they pressed one of the answer buttons. \par
\item \textit{Time till first moves}: Furthermore, we assessed the time until the first move of the mouse in the task by counting the frames from clicking the start button until the first mouse move was detected.\par
\item \textit{Characteristics of the Movement Path}: First, we calculated the travelled distance per path, by summing up the euclidean distance between each sample. Second, we analyzed certain characteristics of the movement path itself. To do so, we first normalized the mouse positions, meaning that the origin of every mouse trail is the origin of the coordinate system. Since at 50\% of the trials, the position of the buttons was flipped, we mirrored the paths on the X-axis of the screen. Next, we calculated the optimal line from the first sample to the last sample of the mouse positions. After that, we calculated the distance of each sample to this line, representing the deviation from the optimal path to the final destination. Building on these normalized deviations from the optimal travel path, we calculated the skew and kurtosis of these distances following the suggestions of prior work about the relationship between traits and the experience of anxiety as well as their personality traits \cite{Dechant2022DontTask}. Moreover, we calculated the area between the optimal line and the absolute distance to this line for each sample. Finally, we analyzed the maximum distance from the optimal line as a complimentary measurement. \par
\end{itemize}

\subsection{Expected behaviour}
Building on the large application of the mouse trajectory in adjunct research, we derive the following expectations about the relationship between the analyzed characteristics of the user's behaviour in the task:

\subsubsection{Time Measurements} 
As previously discussed, time has played a central role in detecting whether an individual may show trustful behaviour. In particular, prior work shows that individuals, who trusted a system, may be faster in their decisions \cite{HarrisonMcKnight2001TrustTime,Hu2016Real-TimeFoundation.}. Prior work on human interaction emphasizes the critical role of thinking time within the context of trust \citet{Efendic2020SlowPredictions}, suggesting that longer response times of a human may be perceived as more trust-inducing. However, the opposite was observed when users had to interact with a suggestion of an algorithm, showing that a shorter calculation time of the system may mean the user trusts the system. Building on these observations, we hypothesise that similar effects may appear in our system: if a user trusts the system, they will perform faster since they don't have to spend time thinking about the content of the tweet being offensive or not. If the user does \textit{not} trust the system, they will think longer about the question, resulting in more time spent in the trial and a longer waiting phase at the beginning of the task. 

\subsubsection{Speed Measurements}
Building on prior work \cite{Cho2006OnlineHesitation}, we hypothesise that a shorter time in the trial will result in faster general mouse movement for individuals scoring high in trust and confidence. Due to a potentially shorter path but also a shorter time we further hypothesise that the maximum speed per trial will be higher.

\subsubsection{Distance Measurements}
As prior work in related fields suggests, hesitation can be caused by the lack of trust, for example, in the context of online shopping \cite{Cho2006OnlineHesitation,Jarvenpaa2000ConsumerStore}, where participants hold back purchase actions due to an increased perceived risk at certain online shops. Previous research in the context of trust suggested other adjunct measurements to the concept of hesitation, like how often individuals change their decision due to the suggestions of an AI \cite{Vereschak2021HowMethodologies}. Additionally, mouse trajectory was successfully applied as a measurement to detect whether individuals were hesitant about their decisions \cite{Zushi2015AnalysisProblems}. Building on these results, we hypothesise that individuals with elevated levels of trust in the AI's recommendation may show a shorter, more direct path towards their decision or the AI's suggestion. Therefore, the maximal distance from the most direct path will be shorter for people experiencing more trust and confidence in their own decision.

\subsubsection{Path Shape Measurements}
We hypothesise that individuals scoring low in trust may express this through a more curvy path. Therefore the skewness and kurtosis of the path may be affected by a lack of trust or confidence. On the contrary, this means that individuals high in trust or confidence will also have a low skew and kurtosis. The area between the movement path and the direct line may be smaller for individuals high in trust or confidence. 

In summary, we hypothesise that individuals with elevated trust will spend less time on the task, start their mouse movement quicker and have a straighter path toward the suggested button. However, although the literature suggests a similar behaviour, we hypothesise that trustful users may begin the task earlier than confident users, since individuals trusting the system may not need to think about which answer they give and "blindly" trust the AI's suggestions.

\subsection{Data analysis}
In total, we recruited 182 participants at the beginning of the analysis. First, we excluded participants who did not finish the experiment, e.g., due to technical difficulties or incomplete data, or showed negligent behaviour, such as rushing through the questionnaires and selecting random answers (n = 34). Since prior work emphasizes the influence of gender on the manifestation of trust we encoded gender (1 = male, female = 0) and removed non-binary people from the analysis (n = 4). Finally, this resulted in \textit{5840 samples} of 146 participants for the following data analysis. Furthermore, we encoded the level of presented accuracy of the AI (1 = 90\% accuracy, 0 = 60\% accuracy) since we treated these levels as a factor (e.g. "high performing AI" versus "low performing AI"). 
To estimate whether there is a relationship between self-reported situational trust in AI as well as confidence in one’s decision and behavioural markers, we conducted multiple regression analysis using IBM SPSS Statistics 27. We performed one regression for each prior introduced marker (resulting in a total of nine multiple regressions). In block one of the analyses, we entered the control variables: gender, age, and indicated AI accuracy. In block two, we entered the self-reported situational trust (in the AI) as well as the self-reported confidence (in one’s own decision) scores. In block three, we added the interaction term trust*confidence to check whether self-reported confidence moderated the relationship between situational trust in the AI and each marker that has been used as the dependent variable. In the following section, we describe each model of these multiple regression analyses.
\section{Results}

\subsection{Relationship between mean subjective after trial ratings and trait trust and trait confidence questionnaires}
As previously discussed, there is an ongoing debate whether the usage of a single-item rating of trust and confidence may not be enough to capture full construct of trust \cite{Brzowski2019TrustReview,Chita-Tegmark2021CanMeasure,Vereschak2021HowMethodologies}. Therefore, we analyzed the relationship between each self-report rating of experienced trust and confidence (after each trial) and validated questionnaires. First, we calculated the mean score of trust and confidence per participant. After that we used a Pearson's Correlation to investigate the relationship between situational trust and confidence measurements and dispositional trust in others, trust in automation, and confidence in their decision. Results indicate a positive relationship between the mean per participant of the two of experienced trust and experienced confidence after each trial (Pearson's r = .204,  p = .013). Further, we found a negative correlation between the mean situational trust and the trust in automation scale of \citet{Merritt2013ISystem} (Pearson's r = -.212,  p = .010). Overall, these results suggest that our measurement of trust is related to the measurement of trust in automation while the confidence measurement is not related to confidence in their decision. \autoref{table_demographics} summarizes the demographic information and the results of the subjective measurements of this sample.

\begin{table}[ht]
\resizebox{\columnwidth}{!}{
\begin{tabular}{rlllllll}
\hline
Variable                                    & Categories           & n   & \%   & M      & SD     & Min & Max \\ \hline
Age                                         &                      & 146 &      & 41.164 & 13.556 & 20  & 70  \\ \hline
Gender                                      & Woman                & 63  & 43.1 &        &        &     &     \\
                                            & Male                 & 83  & 58.0 &        &        &     &     \\
                                            & Non-Binary           & 4   & 2.7  &        &        &     &     \\
                                            & Prefer Not To Answer & 0   &      &        &        &     &     \\ \hline
Wrightsman et al. Dispositional Trust Scale &                      & 146 &      & -8.329 & 7.695  & -27 & 16  \\ \hline
Merritt et al. Trust in Automation Scale    &                      & 146 &      & 17.315 & 3.383  & 10  & 26  \\ \hline
Rosenberg et al. Self Esteem Scale          &                      & 146 &      & 25.986 & 4.07   & 16  & 38 
\end{tabular}}
\caption{Summary of the demographic information, Merritt et al. Trust in Automation Scale, Wrightsman et al. Dispositional Trust Scale, and self-esteem scale.}
\Description{Summary of all demographic information as well as all questionnaire recordings; the left column describes the measurements, then the categories, the percentage, the mean, the standard deviation, the minimum and the maximum of the found values.}
\label{table_demographics}
\end{table}

\subsection{Time Measurements}

\subsubsection{Time spent in trials}
The resulting model for this marker predicted only 3\% of the variance of time spent in trials. 
Participants were likely to spend significantly more time in the trials when they have self-reported that they did not trust the assessment of the AI (\(\beta\)=-.07, t=-5.2, p<.001). Further, they were likely to spend more time in trials when they had low confidence in their own decision (\(\beta\)=-.13, t=-10.05, p<.001). 
There was a significant interaction effect (moderation) of situational trust and self-reported confidence on the time spent in trials (\(\beta\)=.20, t=2.58, p<.010). This means that individuals scoring low in confidence were likely to take more time than individuals scoring low in trust. However, high scoring individuals of both groups result in similarly low times, when interacting with a highly accurate AI. However, when interacting with an AI with a low accuracy, individuals with low trust scores were likely to be very fast. However, AI accuracy had no influence on the relationship between confidence and time which means, that individuals scoring low in confidence in both condition were likely to spend more time in the task than individuals scoring high in confidence. We further observed that one of our control variables (age) predicted time spent in trials (\(\beta\)=.07, t=5.58, p<.001), with older participants being likely to be slower. \autoref{fig_Results} visualizes these results. 
\begin{table}[ht]
\resizebox{\columnwidth}{!}{
\begin{tabular}{l|l|l|l|l|l}
\textbf{Dependent Variable: Time in Trial} & \textbf{B}      & \textbf{Std. Error} & \textbf{beta}  & \textbf{t}       & \textbf{p}    \\ \hline
Age                               & \textbf{.036}   & \textbf{.007}       & \textbf{.072}  & \textbf{5.577}   & \textbf{.000} \\
Gender                                     & -.207           & .161                & -.017          & -1.281           & .200          \\
AI Confidence                              & .095            & .161                & .008           & .590             & .556          \\
Subjective Trust                           & \textbf{-1.385} & \textbf{.266}       & \textbf{-.071} & \textbf{-5.202}  & \textbf{.000} \\
Subjective Confidence                      & \textbf{-4.227} & \textbf{.421}       & \textbf{-.132} & \textbf{-10.051} & \textbf{.000} \\
Subjective Trust x Subjective Confidence   & \textbf{3.689}  & \textbf{1.428}      & \textbf{.195}  & \textbf{2.584}   & \textbf{.010}
\end{tabular}}
\caption{Summary of the model about the relationship between spent time in trial and the subjective measurements of trust and confidence. Significant results are bold highlighted.}
\label{table_timeInTrial}
\end{table}

\subsubsection{Time before first cursor movement} 
The model predicted 1.6\% of variance for the number of samples before they first moved the cursor. \autoref{fig_Results} visualizes these results.  
It took participants significantly longer to move the mouse cursor for the first time in trials in which they self-reported a low confidence in their own assessment (\(\beta\)=-.118, t=-8.91, p<.001). Situational trust in the AI did not predict how long it took participants to move their cursor (\(\beta\)=-.02, t=-1.50, p<.133). There was no significant interaction effect between self-reported trust and confidence in the participants’ own judgment on the time that it took participants to move the cursor for the first time (\(\beta\)=.08, t=1.05, p<.295). Older participants tended to take longer to move the mouse (\(\beta\)=.03, t=2.23, p<.026) and surprisingly, participants took longer to move the mouse in trials in which the accuracy of the AI had been indicated to be higher (\(\beta\)=.03, t=2.02, p<.044).
\begin{table}[ht]
\resizebox{\columnwidth}{!}{
\begin{tabular}{l|l|l|l|l|l}
\textbf{\begin{tabular}[c]{@{}l@{}}Dependent Variable: \\ Sample count before first move\end{tabular}} & \textbf{B}       & \textbf{Std. Error} & \textbf{beta}  & \textbf{t}      & \textbf{p}    \\ \hline
Age                                                                                                    & \textbf{.241}    & \textbf{.108}       & \textbf{.29}   & \textbf{2.232}  & \textbf{.026} \\
Gender                                                                                                 & -4.961           & 2.674               & -.024          & -1.855          & .064          \\
AI Confidence                                                                                          & \textbf{5.36}    & \textbf{2.661}      & \textbf{.026}  & \textbf{2.017}  & \textbf{.044} \\
Subjective Trust                                                                                       & -6.673           & 4.437               & -.021          & -1.504          & .133          \\
Subjective Confidence                                                                                  & \textbf{-62.449} & \textbf{7.007}      & \textbf{-.118} & \textbf{-8.912} & \textbf{.000} \\
Subjective Trust x Subjective Confidence                                                               & 24.920           & 23.802              & .080           & 1.047           & .295         
\end{tabular}}
\caption{Relationship between samples count before the first mouse move and the subjective measurements of trust and confidence. Significant results are bold highlighted.}
\label{table_samplesBeforeMove}
\end{table}

\subsection{Speed Measurements}
\subsubsection{Maximum speed}
This model only explained 0.8\% of variance for maximum speed.
The maximum speed of mouse movement was predicted by situational trust in the AI (\(\beta\)=.05, t=3.34, p<.001), meaning that people who trusted the AI more were likely to be faster. There was no significant effect of confidence in one’s own judgment (\(\beta\)=.01, t=.30, p<.765) and confidence did not moderate the relationship between trust in the AI and maximum speed (\(\beta\)=.09, t=1.17, p<.242).
This marker was further predicted by participant age (\(\beta\)=-.08, t=-5.83, p<.001) and gender (\(\beta\)=-.03, t=-2.38, p<.017), with older participants being more likely to reach less speed and women being likely to reach a higher maximum speed than men. \autoref{fig_Results} visualizes these results. 
\begin{table}[ht]
\resizebox{\columnwidth}{!}{
\begin{tabular}{l|l|l|l|l|l}

\textbf{\begin{tabular}[c]{@{}l@{}}Dependent Variable: \\ Maximum Speed\end{tabular}} & \textbf{B}      & \textbf{Std. Error} & \textbf{beta}  & \textbf{t}      & \textbf{p}    \\ \hline
Age                                                                                   & \textbf{-.107}  & \textbf{.018}       & \textbf{-.075} & \textbf{-5.825} & \textbf{.000} \\
Gender                                                                                & \textbf{-1.088} & \textbf{.457}       & \textbf{-.031} & \textbf{-2.379} & \textbf{.017} \\
AI Confidence                                                                         & .391            & .455                & .911           & .860            & .390          \\
Subjective Trust                                                                      & \textbf{2.551}  & \textbf{.764}       & \textbf{.046}  & \textbf{.341}   & \textbf{.001} \\
Subjective Confidence                                                                 & .361            & 1.206               & .004           & .299            & .765          \\
Subjective Trust x Subjective Confidence                                              & 4.791           & 4.097               & .089           & 1.170           & .242         
\end{tabular}}
\caption{Relationship between maximum speed and the subjective measurements of trust and confidence. Significant results are bold highlighted.}
\label{table_maxSpeed}
\end{table}

\subsubsection{Mean speed}
The model predicted 8\% of the variance for mean cursor movement speed.
For average speed we find that participants overall were likely to move faster if they self-reported high trust in the AI (\(\beta\)=.12, t=8.90, p<.001) and high confidence (\(\beta\)=.17, t=13.27, p<.001) in their own decision. There was no significant interaction of self-reported trust in the AI and self-reported confidence in one’s own decision on mean speed (\(\beta\)=-.13, t=-1.82, p<.069).
Two of our control variables - gender (\(\beta\)=.03, t=1.98, p<.048) and age (\(\beta\)=-.18, t=-13.72, p<.001) - further had an influence on speed. Younger participants and men were more likely to move the cursor faster on average.
\begin{table}[ht]
\resizebox{\columnwidth}{!}{
\begin{tabular}{l|l|l|l|l|l}
\textbf{\begin{tabular}[c]{@{}l@{}}Dependent Variable: \\ Mean Speed\end{tabular}} & \textbf{B}     & \textbf{Std. Error} & \textbf{beta}  & \textbf{t}       & \textbf{p}    \\ \hline
Age                                                                                & \textbf{-.005} & \textbf{.000}       & \textbf{-.176} & \textbf{-13.723} & \textbf{.000} \\
Gender                                                                             & \textbf{.17}   & \textbf{.009}       & \textbf{.205}  & \textbf{1.977}   & \textbf{.048} \\
AI Confidence                                                                      & .013           & .009                & .020           & .1537            & .124          \\
Subjective Trust                                                                   & \textbf{.126}  & \textbf{.014}       & \textbf{.118}  & \textbf{8.903}   & \textbf{.000} \\
Subjective Confidence                                                              & \textbf{.298}  & \textbf{.022}       & \textbf{.169}  & \textbf{13.273}  & \textbf{.000} \\
Subjective Trust x Subjective Confidence                                           & -.139          & .076                & -.134          & -1.821           & .069         
\end{tabular}}
\caption{Relationship between mean speed and the subjective measurements of trust and confidence. Significant results are bold highlighted.}
\label{table_meanSpeed}
\end{table}

\subsection{Distance Measurements}
\subsubsection{Travelled distance of the cursor}
Only 0.9\% of the variance of travelled distance was explained by the model.
Participants moved their cursor more when they self-reported a low trust in the AI assessment (\(\beta\)=-.05, t=-3.79, p<.001) as well as low confidence in their own judgment (\(\beta\)=-.05, t=-3.48, p<.001). There was no interaction effect of self-reported confidence in one’s own decision and trust in the AI assessment on the distance that the cursor had been moved (\(\beta\)=.09, t=1.22, p<.225).
Further, our control variable gender had a significant effect on travelled cursor distance, as women tended to move the mouse more than men (\(\beta\)=-.06, t=-4.37, p<.001). 
\begin{table}[ht]
\resizebox{\columnwidth}{!}{
\begin{tabular}{l|l|l|l|l|l}
\textbf{\begin{tabular}[c]{@{}l@{}}Dependent Variable: \\ Travelled Distance\end{tabular}} & \textbf{B}        & \textbf{Std. Error} & \textbf{beta}  & \textbf{t}      & \textbf{p}    \\ \hline
Age                                                                                        & 1.419             & 1.101               & .017           & 1.288           & .198          \\
Gender                                                                                     & \textbf{-119.316} & \textbf{27.330}     & \textbf{-.57}  & \textbf{-4.366} & \textbf{.000} \\
AI Confidence                                                                              & -41.271           & 27.197              & -.202          & -1.517          & .129          \\
Subjective Trust                                                                           & \textbf{-172.638} & \textbf{45.563}     & \textbf{-0.52} & \textbf{-3.789} & \textbf{.000} \\
Subjective Confidence                                                                      & \textbf{-250.211} & \textbf{71.965}     & \textbf{-.046} & \textbf{-3.477} & \textbf{.001} \\
Subjective Trust x Subjective Confidence                                                   & 296.866           & 244.433             & .093           & 1.215           & .225         
\end{tabular}}
\caption{Summary of the model about the relationship between travelled distance and the subjective measurements of trust and confidence. Significant results are bold highlighted.}
\label{table_travelledDistance}
\end{table}

\subsubsection{Maximal distance from the most direct path}
Less than 0.1\% of the variance of travelled distance was explained by the model.
Neither trust in the AI (\(\beta\)=.01, t=.71, p<.479) nor confidence in one’s own decision (\(\beta\)=.01, t=1.01, p<.313) predicted how far participants strayed from the most direct mouse path (to the button of the decision they were making). There was no interaction effect of trust and confidence (\(\beta\)=.05, t=.65, p<.514). Women were likely to take a longer path to the button when making their decision (\(\beta\)=-.04, t=-2.67, p<.008).
\begin{table}[ht]
\resizebox{\columnwidth}{!}{
\begin{tabular}{l|l|l|l|l|l}
\textbf{\begin{tabular}[c]{@{}l@{}}Dependent Variable: \\ Maximum Distance from Direct Path\end{tabular}} & \textbf{B}       & \textbf{Std. Error} & \textbf{beta}  & \textbf{t}      & \textbf{p}    \\ \hline
Age                                                                                                       & -.149            & .318                & -.006          & -4.68           & .640          \\
Gender                                                                                                    & \textbf{-21.119} & \textbf{7.902}      & \textbf{-.035} & \textbf{-2.673} & \textbf{.008} \\
AI Confidence                                                                                             & -4.004           & 7.863               & -.007          & -.509           & .611          \\
Subjective Trust                                                                                          & 9.351            & 13.208              & .010           & .708            & .479          \\
Subjective Confidence                                                                                     & 21.050           & 20.861              & .013           & 1.009           & .313          \\
Subjective Trust x Subjective Confidence                                                                  & 46.208           & 70.861              & .050           & .652            & .514         
\end{tabular}}
\caption{Relationship between maximal distance to the direct path and the subjective measurements of trust and confidence. Significant results are bold highlighted.}
\label{table_maxDistance}
\end{table}

\subsection{Path Shape Measurements}
\subsubsection{Kurtosis and Skew}
The model explained less than 0.1\% of variance in the mouse path kurtosis.
The centre of the curve for the mouse movement (kurtosis) was higher for participants with low confidence in their own decision (\(\beta\)=-.04, t=-3.09, p<.002). There was no effect of trust in the AI (\(\beta\)=.01, t=.33, p<.743) and no interaction effect (\(\beta\)=-.07, t=-.92, p<.355). 
The kurtosis was further likely to be higher for men than for women (\(\beta\)=.04, t=2.94, p<.003). 
Similar to that the model about path skewness predicted less than 0.1\% of variance in mouse path skew.
There were no effects found for trust (\(\beta\)=.02, t=1.18, p<.237), confidence (\(\beta\)=-.01, t=-.78, p<.439), nor their interaction (\(\beta\)=-.08, t=-.99, p<.321) on skew. 
For men the curve was more skewed to the left than it was for women (\(\beta\)=.04, t=2.74, p<.006). 
\begin{table}[ht]
\resizebox{\columnwidth}{!}{
\begin{tabular}{l|l|l|l|l|l}
\textbf{\begin{tabular}[c]{@{}l@{}}Dependent Variable: \\ Path Kurtosis\end{tabular}} & \textbf{B}      & \textbf{Std. Error} & \textbf{beta}  & \textbf{t}      & \textbf{p}    \\ \hline
Age                                                                                   & -.017           & .019                & -.012          & -.927           & .354          \\
Gender                                                                                & \textbf{1.352}  & \textbf{.461}       & \textbf{.038}  & \textbf{2.935}  & \textbf{.003} \\
AI Confidence                                                                         & .162            & .458                & .005           & .354            & .723          \\
Subjective Trust                                                                      & .252            & .769                & .005           & .328            & .743          \\
Subjective Confidence                                                                 & \textbf{-3.752} & \textbf{1.215}      & \textbf{-.041} & \textbf{-3.087} & \textbf{.002} \\
Subjective Trust x Subjective Confidence                                              & -3.816          & 4.128               & -.071          & -.924           & .355         
\end{tabular}}
\caption{Relationship between samples count before the path kurtosis and the subjective measurements of trust and confidence. Significant results are bold highlighted.}
\label{table_kurtosis}
\end{table}

\begin{table}[ht]
\resizebox{\columnwidth}{!}{
\begin{tabular}{l|l|l|l|l|l}

\textbf{\begin{tabular}[c]{@{}l@{}}Dependent Variable: \\ Path Skew\end{tabular}} & \textbf{B}    & \textbf{Std. Error} & \textbf{beta} & \textbf{t}     & \textbf{p}    \\ \hline
Age                                                                               & .002          & .002                & .020          & 1.519          & .129          \\
Gender                                                                            & \textbf{.107} & \textbf{.039}       & \textbf{.036} & \textbf{2.744} & \textbf{.006} \\
AI Confidence                                                                     & -.017         & .39                 & -.006         & -.432          & .665          \\
Subjective Trust                                                                  & .077          & .065                & .016          & 1.182          & .237          \\
Subjective Confidence                                                             & -.080         & .103                & -.010         & -.775          & .439          \\
Subjective Trust x Subjective Confidence                                          & -.347         & .350                & -.076         & -.992          & .321         
\end{tabular}}
\caption{Relationship between skewness of the path and the subjective measurements of trust and confidence. Significant results are bold highlighted.}
\label{table_skew}
\end{table}

\subsubsection{Area between the mouse path and the most direct path}
Less than 0.1\% of the variance of travelled distance was explained by the model.
Neither trust in the AI (\(\beta\)=.01, t=.50, p<.621) nor confidence in one’s own decision (\(\beta\)=.02, t=1.51, p<.131) predicted the normalized area of the mouse path. There was no interaction effect of trust and confidence (\(\beta\)=.04, t=.54, p<.592). Women were likely to use a larger area for their mouse path (\(\beta\)=-.05, t=-.3.67, p<.001).
\begin{table}[ht]
\resizebox{\columnwidth}{!}{
\begin{tabular}{l|l|l|l|l|l}
\textbf{\begin{tabular}[c]{@{}l@{}}Dependent Variable: \\ Maximum Distance from Path Volume\end{tabular}} & \textbf{B}           & \textbf{Std. Error} & \textbf{beta}  & \textbf{t}      & \textbf{p}    \\ \hline
Age                                                                                                       & -576.322             & 1136.167            & -.007          & -.507           & .612          \\
Gender                                                                                                    & \textbf{-103371.025} & \textbf{28191.526}  & \textbf{-.048} & \textbf{-3.667} & \textbf{.000} \\
AI Confidence                                                                                             & -14112.15            & 28054.396           & -.007          & -.503           & .615          \\
Subjective Trust                                                                                          & 23323.303            & 47117.219           & .007           & .496            & .621          \\
Subjective Confidence                                                                                     & 112435.613           & 74419.217           & .020           & 1.511           & .131          \\
Subjective Trust x Subjective Confidence                                                                  & -21990.745           & 29241.044           & -,010          & -,752           & .452         
\end{tabular}}
\caption{Relationship between path volume and the subjective measurements of trust and confidence. Significant results are bold highlighted.}
\label{table_volumee}
\end{table}

\begin{figure}[htp]
\includegraphics[width=1\textwidth]{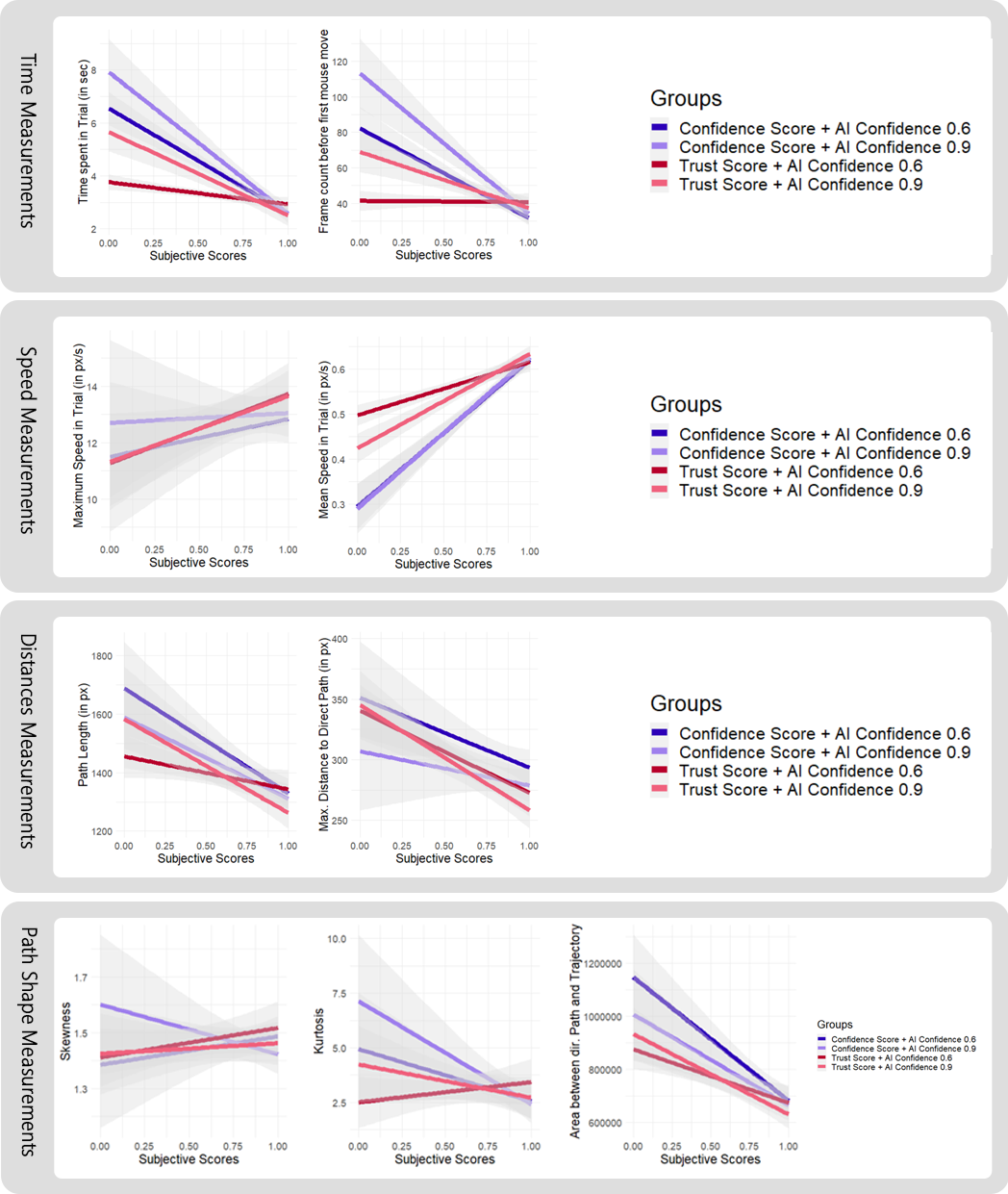}
\centering
\caption{The results grouped by time, speed, distance, and path shape characteristics.}
\Description{The figure shows the results of our study. row one summarizes the time measurements. Both time in trial and samples before first mouse move show that trust and confidence measurements go down, meaning that users who were more confident or trusting the system were also faster and moved their mouse faster. However trust measurements were not significant and almost no slope.

Row 2 shows positive relationship between max speed and mean speed for both trust and confidence while the AI Accuracy has no effect.

Row 3 shows a negative relationship between path length and max distance from direct line to destination for both trust and confidence while the AI Accuracy has no effect.

Row 4 shows the results of Skewness, Kurtosis and Area between direct path and mouse trajectory. Skewness has no clear direction of the relationships. Kurtosis and Area show a negative relationship for both trust and confidence while the AI Accuracy has no effect.
}
\label{fig_Results}
\end{figure}

\autoref{fig_Results} visualize the results of our analysis. 

\section{Discussion}

\subsection{Summary of Results}
We investigated the relationship between subjective ratings of trust, confidence, and the user's mouse movement besides the influence of age and gender-identity. We found the following significant interactions.\autoref{table_summary} highlights all significant relationships between the measurements, the subjective measurements of trust and confidence as well as the percentage of explained variance.
\begin{table}[ht]
\resizebox{\columnwidth}{!}{
\begin{tabular}{l|l|lll|lll}
                                 & \% of explainable variance & Age & Gender & AI accuracy & Score Trust & Score Confidence & \begin{tabular}[c]{@{}l@{}}Score Trust x\\ Score Confidence\end{tabular} \\ \hline
Time in Task                     & 3\%                        &  \checkmark   &  \checkmark      &             &  \checkmark           &  \checkmark                &  \checkmark                                                                        \\
Samples before first move        & 1.6 \%                     &  \checkmark   &        &  \checkmark           &             &  \checkmark                &                                                                          \\ \hline
Maximum Speed                    & 0.8 \%                     &  \checkmark   &  \checkmark      &             &             &                  &                                                                          \\
Mean Speed                       & 8 \%                       &  \checkmark   &  \checkmark      &             &  \checkmark           &  \checkmark                &                                                                          \\ \hline
Travelled Distance               & 0.9 \%                     &     &  \checkmark      &             &  \checkmark           &  \checkmark                &                                                                          \\
Max. Distance of the Direct Path & 0.1 \%                     &     &  \checkmark      &             &             &                  &                                                                          \\ \hline
Kurtosis                         & \textless 0.1 \%           &     & \checkmark      &             &             & \checkmark                &                                                                          \\
Skew                             & \textless 0.1 \%           &     & \checkmark      &             &             &                  &                                                                          \\
Path Volume                      & \textless 0.1 \%           &     & \checkmark      &             &             &                  &                                                                         
\end{tabular}}
\caption{Summary of all significant relationships between the measurements, the subjective measurements and the found explanation of variance per model.}
\Description{Overview of which measurement revealed a significant relationship between subjective trust, subjective confidence, the interaction of both as well the percentage of explained variance per model.}
\label{table_summary}
\end{table}

\subsubsection{Trust}
We show that an increased self-reported trust led to a shorter overall time spent in the task, yet we found no significant influence of self-reported trust on the time span before the first move of the mouse. We observe an interaction effect between trust, confidence and time spent in the task (see \autoref{fig_Results}). Further, we show that higher trust may result in an overall shorter path. Increased levels of trust led to increased average mouse movement speed. However, we found no significant relationships between trust and maximum speed per trial, distance between mouse path and most direct path, nor skewness and kurtosis of the trajectory.

\subsubsection{Confidence}
Similar to trust, we found a significant relationship between self-reported confidence and time spent in trial, resulting in a reduced time in the task for trustful individuals. However, our results show that high decision-confidence led to a decreased time before the first mouse movement. Similar to trust, an increased level of confidence in their decision led to a shorter path length and a faster mean speed. We found no significant relationship between confidence and maximum speed, volume of the mouse path, maximum distance from the optimal path towards the button, and the skewness of the mouse path. However, we found a significant relationship between kurtosis and self-reported confidence.

\subsection{Explanation of Findings}
Our results suggest that time, path length and mean speed, are affected by the experience of trust in the system. Prior work suggests similar results about time being related to trustful behaviour \cite{Pu2007Trust-inspiringSystems,Yuksel2017BrainsInteractions}: individuals who were unsure about their decisions spent more time in the task than individuals following the suggestions of an AI-enhanced system. Some researchers argue that trust is a tool for individuals to reduce the complexity of an interaction \cite{Gilbert1998TowardsTrust}. This means that making a decision becomes easier due to the experience of trust and therefore the individual may not need to expend cognitive resources in order to process a given situation. Instead they can rely on the suggestions of a trusted entity \cite{Vereschak2021HowMethodologies}. Similar conceptualizations were found in the context of trust in professional organisations \cite{Gilbert2005ImpersonalDynamics}. As prior work suggests, mental models about the system play a critical role which may explain the faster movements \cite{Hastie2017TrustInterfaces}. Further, our results regarding path length are in line with prior work about trust in security systems, suggesting that the travelled mouse distance may be useful for identifying a trustful user \cite{Pu2007Trust-inspiringSystems}.
While these results are promising there are still challenging characteristics of trust: as discussed, trust is an attitude, which may express in unique ways. As shown, while there are some specific aspects of the mouse trajectory, which may be seen as an expression of trust \cite{Yi2020IdentificationEmotions}, the explained variance may not be sufficient. This could be explained by trust being seen as an attitude \cite{Vereschak2021HowMethodologies} unlike other emotional states such as fear or anxiety \cite{Salkovskis1991TheAccount}. Furthermore, trustful behaviour may be seen as a "reaction" to incoming signals \cite{Riedl2012TheImaging.,Adolphs2002TrustBrain} and more subtle than other mental states or personality traits, which shape the user's behaviour more drastically than the experience of situational trust.

While prior work mostly focused on trust as a single term, our results show large overlaps with related constructs, such as confidence in their decision. As previously discussed, the overlap between these two constructs may be explained through the underlying relationship between trust and confidence: while trust focuses on the interaction with someone else and the vulnerability towards the \textit{other's decision}, confidence only focuses on the self and the vulnerability towards the overall situation \cite{Vereschak2021HowMethodologies}. Some researchers argue that trust can be seen as \textit{confidence in the other's decision} \cite{Luhmann1988Trust:Relations}. Our results confirm the close relationship between trust and confidence, and show the urgent need of precise measurements in order to properly assess trustful behaviour distinct from or in combination with related constructs such as confidence.

\subsection{Implications for Design}
Our work shows the challenging aspects of assessing the user's trust in an AI enhanced system and shows that behavioural data gained from a low-cost approach like the assessment of the mouse trajectory may be an additional tool to assess the level of trust. On one hand researchers have an easy accessible tool, even for remote assessment, at their hand yet our work shows potential shortcomings of this approach: as shown, the multiple regression analyses explained only a small portion of the variance in each analysis. This could also come from the potential limited experience of vulnerability of the users in the presented task, which plays an important role in the experience of trust \cite{Vereschak2021HowMethodologies}. Furthermore our work shows the challenges of the attitude characteristic of trust, resulting in subtle but hard to measure behavioural expression of trust.
Researchers who wish to apply this approach need to balance the accuracy of the measurement by applying additional measurements to gain more insights about the user's trust in a system while respecting the user's comfort. For example, some researchers suggest harnessing not only behavioural data but also cognitive information or markers which may reveal more information about the current cognitive load and state of the user \cite{Castelfranchi2010_Socio,Brzowski2019TrustReview}. Combining different sources of behavioural data, will allow researchers to increase the quality of trust assessment.
Our results emphasize that researchers must be aware of correlations between related concepts, like trust and confidence in own decisions. As the definition of trust shows, confidence may be experienced when the user does not need to rely on another entity\cite{Lee2004TrustReliance,Hosmer1995Trust:Ethics}. In our results, trust and confidence manifested in a similar way and we may, thus, be vulnerable to interpreting confidence as trustful behaviour. Without any additional measurement, researchers will face challenges to differentiate between closely related concepts. As our work, in combination with prior work, suggests, the combination of self-report as well as objective data should be used in order to better understand which mental state may affect the user's behaviour. Researchers in the mental health community face similar challenges and introduced the concept of evidence based assessment \cite{Hunsley2007Evidence-BasedAssessment,Connors2015Evidence-basedHealth}, which combines subjective user inputs with empirically measured evidence, like digital behavioural markers. We emphasize that researchers should follow a similar approach when assessing trust or trustful behaviour. As shown in our experiment, observed behaviour may be caused by several adjunct concepts, such as trust and confidence. 

\subsection{Ethical Implications of the Trust Measurement}
As prior work shows, behavioural measurements such as the mouse movement \cite{Yi2020IdentificationEmotions} or other digital data may reveal private insights about the user, ranging from personal preferences \cite{Mittelstadt2016,Ryan2020InReliability}, mental health concerns \cite{JohannesDechant2021AssessingTask,Freeman2017VirtualDisorders} but also social information such as whether the user may express trust in a system. Additionally, through the integration of more advanced sensors, such as gaze trackers or GPS, researchers gained access to additional data sources, similar to adjunct research projects about detecting confusion \cite{Hosp2021StatesSurgery}, expertise \cite{Hosp2021SoccerMovements}, or even anxiety through gaze information \cite{Dechant2017}. However, this trend inspired an ongoing discussion about the ethical usage of such approaches \cite{Floridi2016,Bruckman2014ResearchHCI,Mittelstadt2016}. When harnessing behavioural data such as mouse movement or the recording of embedded sensor data, we must be aware of potentially problematic aspects of such assessment, including inferring identity, the breach of privacy, and how these approaches may affect users in their social environment. For example, prior work suggests that mouse trajectory not only may reveal information about the user's trust in another entity, human or AI, but also about the personal performance and preference \cite{Floridi2016}. Such highly sensitive data may be misused against the individual, for example as a reason to dismiss a worker due to low performance \cite{Bartels2012}.
The assessment of trust is challenging due to its attitude characteristic. This means, that although the individual may trust a machine or others, it does not mean they will express it through their behaviour. Through the application of evidence based assessment, researchers build a solid foundation to better understand the current level of trust in a system. However, when applying this knowledge, we must ensure the protection of the user's privacy. 

\subsection{Limitations and Future work}
There are several limitations which may be addressed by future work to further strengthen our understanding of how trust expresses through behavioural markers. 
First, our research focused on mouse trajectory and analysed its relationship with trust and confidence. Future work may include additional behavioural measurements such as gaze data \cite{Barbato2020TheOrienting}, since prior work suggests a close relationship between gaze and the user's mouse movement. This additional verification step may strengthen the reliability of mouse trajectory. Furthermore future work should look into assessing trust in the field by repeating this study with a more applied interaction task to better understand the influence of user interface components on observed trust and confidence behavioural markers. With the rise of alternative ways to interact with a growing diversity of information system types, such as touch, we suggest that future work may look into other inputs, such as touch or even full body and hand gestures. Besides input devices, future work may also investigate different platforms, such as mobile devices or immersive environments in order to fully understand which aspects of the platform and the input modality can be used as a behavioural measurement for trust. Furthermore, we had no information about the used mouse and hardware setup, which may also bias results, for example by an increased DPI-setting of the mouse which may result in a faster mouse cursor. Future work should analyze the influence of certain hardware components, such as the mouse, but also the used software, e.g. the internet-browser, to better understand these influences on the potential usage of mouse trajectory as a measurement for situational trust.
Second, the surrounding environment: in our task, users were placed in a low risk environment and were instructed. However, as the literature suggests, stress and other environmental factors may affect user behaviour. AI assistant systems will often be applied in dangerous environments, where trusting the system may be crucial. Therefore, future research should investigate the role of the environment on the suggested behavioural markers for trustful behaviour. 
Third, our work focused on the interaction with an artificial intelligence. However, trust and confidence are essential components for interactions with others. Future work should look into whether the presence of a human or a simulated AI affects the manifestation of trust and confidence through mouse trajectory. This would also help to better understand the role of the AI's performance on the observed relationships. Other studies show that a poor performance of the AI may lead to a decrease of trust. However, our results show a low influence of the AI performance on the marker. Through a task adapted from a real world setup, such as the context of decision making in a high risk environment \cite{Solberg2022AAids}, future researchers may be able to better understand the role of AI performance on the expression of trustful and confidence behaviour.   

\section{Conclusion}
Trust is a core component for successful social interactions in our daily life with other humans, but also with digital solutions and tools. In recent years, researchers tried to look for ways to increase the user's trust in artificial intelligent systems. On the one hand, prior work suggested expensive and potentially invasive sensors to assess the level of trust. On the other hand, some researchers suggest the mouse trajectory as a handy and cheap tool. Yet, the relationship between mouse movement, trust, and related constructs like confidence, are not fully understood. Without that knowledge we risk to collect noisy or distorted data, which cannot accurately assess trust. Therefore, we conducted an online study in which participants had to work together with two versions of a simulated AI and had to decide whether presented tweets were offensive or not. We show that time, mean movement speed of the mouse and path length have the strongest relationship with self-report measures of trust, yet the models only explain a small amount of the variance in the data. We suggest the use of mouse trajectory as a measure for trustful behaviour, but also emphasize the urge to combine multiple empirical measurements with self-report scales to fully understand whether user behaviours, such as their mouse movement, truly expresses trust.

\begin{acks}
Removed for review
\end{acks}

\bibliographystyle{ACM-Reference-Format}
\bibliography{md_References}


\begin{thebibliography}{112}


\ifx \showCODEN    \undefined \def \showCODEN     #1{\unskip}     \fi
\ifx \showDOI      \undefined \def \showDOI       #1{#1}\fi
\ifx \showISBNx    \undefined \def \showISBNx     #1{\unskip}     \fi
\ifx \showISBNxiii \undefined \def \showISBNxiii  #1{\unskip}     \fi
\ifx \showISSN     \undefined \def \showISSN      #1{\unskip}     \fi
\ifx \showLCCN     \undefined \def \showLCCN      #1{\unskip}     \fi
\ifx \shownote     \undefined \def \shownote      #1{#1}          \fi
\ifx \showarticletitle \undefined \def \showarticletitle #1{#1}   \fi
\ifx \showURL      \undefined \def \showURL       {\relax}        \fi
\providecommand\bibfield[2]{#2}
\providecommand\bibinfo[2]{#2}
\providecommand\natexlab[1]{#1}
\providecommand\showeprint[2][]{arXiv:#2}

\bibitem[Adolphs(2002)]%
        {Adolphs2002TrustBrain}
\bibfield{author}{\bibinfo{person}{Ralph Adolphs}.}
  \bibinfo{year}{2002}\natexlab{}.
\newblock \showarticletitle{{Trust in the brain}}.
\newblock \bibinfo{journal}{\emph{Nature Neuroscience}} \bibinfo{volume}{5},
  \bibinfo{number}{3} (\bibinfo{date}{3} \bibinfo{year}{2002}),
  \bibinfo{pages}{192--193}.
\newblock
\showISSN{1097-6256}
\urldef\tempurl%
\url{https://doi.org/10.1038/nn0302-192}
\showDOI{\tempurl}


\bibitem[Ahlan and Ahmad(2014)]%
        {Ahlan2014UserModel}
\bibfield{author}{\bibinfo{person}{Abd~Rahman Ahlan} {and}
  \bibinfo{person}{Barroon~Isma’eel Ahmad}.} \bibinfo{year}{2014}\natexlab{}.
\newblock \showarticletitle{{User Acceptance of Health Information Technology
  (HIT) in Developing Countries: A Conceptual Model}}.
\newblock \bibinfo{journal}{\emph{Procedia Technology}}  \bibinfo{volume}{16}
  (\bibinfo{year}{2014}), \bibinfo{pages}{1287--1296}.
\newblock
\showISSN{22120173}
\urldef\tempurl%
\url{https://doi.org/10.1016/j.protcy.2014.10.145}
\showDOI{\tempurl}


\bibitem[Akash et~al\mbox{.}(2018)]%
        {Akash2018AGSR}
\bibfield{author}{\bibinfo{person}{Kumar Akash}, \bibinfo{person}{Wan-Lin Hu},
  \bibinfo{person}{Neera Jain}, {and} \bibinfo{person}{Tahira Reid}.}
  \bibinfo{year}{2018}\natexlab{}.
\newblock \showarticletitle{{A Classification Model for Sensing Human Trust in
  Machines Using EEG and GSR}}.
\newblock \bibinfo{journal}{\emph{ACM Transactions on Interactive Intelligent
  Systems}} \bibinfo{volume}{8}, \bibinfo{number}{4} (\bibinfo{date}{11}
  \bibinfo{year}{2018}), \bibinfo{pages}{1--20}.
\newblock
\showISSN{2160-6455}
\urldef\tempurl%
\url{https://doi.org/10.1145/3132743}
\showDOI{\tempurl}


\bibitem[Arrow(1974)]%
        {Arrow1974TheOrganization}
\bibfield{author}{\bibinfo{person}{Kenneth~J Arrow}.}
  \bibinfo{year}{1974}\natexlab{}.
\newblock \bibinfo{booktitle}{\emph{{The limits of organization}}}.
\newblock \bibinfo{publisher}{WW Norton {\&} Company}.
\newblock
\showISBNx{978-0393093230}


\bibitem[Arroyo et~al\mbox{.}(2006)]%
        {Arroyo2006UsabilityTracks}
\bibfield{author}{\bibinfo{person}{Ernesto Arroyo}, \bibinfo{person}{Ted
  Selker}, {and} \bibinfo{person}{Willy Wei}.} \bibinfo{year}{2006}\natexlab{}.
\newblock \showarticletitle{{Usability tool for analysis of web designs using
  mouse tracks}}. In \bibinfo{booktitle}{\emph{CHI '06 Extended Abstracts on
  Human Factors in Computing Systems}}. \bibinfo{publisher}{ACM},
  \bibinfo{address}{New York, NY, USA}, \bibinfo{pages}{484--489}.
\newblock
\showISBNx{1595932984}
\urldef\tempurl%
\url{https://doi.org/10.1145/1125451.1125557}
\showDOI{\tempurl}


\bibitem[Barbato et~al\mbox{.}(2020)]%
        {Barbato2020TheOrienting}
\bibfield{author}{\bibinfo{person}{Mariapaola Barbato},
  \bibinfo{person}{Aisha~A. Almulla}, {and} \bibinfo{person}{Andrea Marotta}.}
  \bibinfo{year}{2020}\natexlab{}.
\newblock \showarticletitle{{The Effect of Trust on Gaze-Mediated Attentional
  Orienting}}.
\newblock \bibinfo{journal}{\emph{Frontiers in Psychology}}
  \bibinfo{volume}{11} (\bibinfo{date}{7} \bibinfo{year}{2020}).
\newblock
\showISSN{1664-1078}
\urldef\tempurl%
\url{https://doi.org/10.3389/fpsyg.2020.01554}
\showDOI{\tempurl}


\bibitem[Barredo~Arrieta et~al\mbox{.}(2020)]%
        {BarredoArrieta2020ExplainableAI}
\bibfield{author}{\bibinfo{person}{Alejandro Barredo~Arrieta},
  \bibinfo{person}{Natalia D{\'{i}}az-Rodr{\'{i}}guez}, \bibinfo{person}{Javier
  Del~Ser}, \bibinfo{person}{Adrien Bennetot}, \bibinfo{person}{Siham Tabik},
  \bibinfo{person}{Alberto Barbado}, \bibinfo{person}{Salvador Garcia},
  \bibinfo{person}{Sergio Gil-Lopez}, \bibinfo{person}{Daniel Molina},
  \bibinfo{person}{Richard Benjamins}, \bibinfo{person}{Raja Chatila}, {and}
  \bibinfo{person}{Francisco Herrera}.} \bibinfo{year}{2020}\natexlab{}.
\newblock \showarticletitle{{Explainable Artificial Intelligence (XAI):
  Concepts, taxonomies, opportunities and challenges toward responsible AI}}.
\newblock \bibinfo{journal}{\emph{Information Fusion}}  \bibinfo{volume}{58}
  (\bibinfo{date}{6} \bibinfo{year}{2020}), \bibinfo{pages}{82--115}.
\newblock
\showISSN{15662535}
\urldef\tempurl%
\url{https://doi.org/10.1016/j.inffus.2019.12.012}
\showDOI{\tempurl}


\bibitem[Bartels and Marshall(2012)]%
        {Bartels2012}
\bibfield{author}{\bibinfo{person}{Michael Bartels} {and}
  \bibinfo{person}{Sandra~P. Marshall}.} \bibinfo{year}{2012}\natexlab{}.
\newblock \showarticletitle{{Measuring cognitive workload across different eye
  tracking hardware platforms}}. In \bibinfo{booktitle}{\emph{Proceedings of
  the Symposium on Eye Tracking Research and Applications - ETRA '12}}.
  \bibinfo{pages}{161}.
\newblock
\showISBNx{9781450312219}
\urldef\tempurl%
\url{https://doi.org/10.1145/2168556.2168582}
\showDOI{\tempurl}


\bibitem[Bawack et~al\mbox{.}(2022)]%
        {Bawack2022ArtificialReview}
\bibfield{author}{\bibinfo{person}{Ransome~Epie Bawack},
  \bibinfo{person}{Samuel~Fosso Wamba}, \bibinfo{person}{Kevin Daniel~André
  Carillo}, {and} \bibinfo{person}{Shahriar Akter}.}
  \bibinfo{year}{2022}\natexlab{}.
\newblock \showarticletitle{{Artificial intelligence in E-Commerce: a
  bibliometric study and literature review}}.
\newblock \bibinfo{journal}{\emph{Electronic Markets}} \bibinfo{volume}{32},
  \bibinfo{number}{1} (\bibinfo{date}{3} \bibinfo{year}{2022}),
  \bibinfo{pages}{297--338}.
\newblock
\showISSN{1019-6781}
\urldef\tempurl%
\url{https://doi.org/10.1007/s12525-022-00537-z}
\showDOI{\tempurl}


\bibitem[Bliss and Acton(2003)]%
        {Bliss2003AlarmDriving}
\bibfield{author}{\bibinfo{person}{James~P Bliss} {and}
  \bibinfo{person}{Sarah~A Acton}.} \bibinfo{year}{2003}\natexlab{}.
\newblock \showarticletitle{{Alarm mistrust in automobiles: how collision alarm
  reliability affects driving}}.
\newblock \bibinfo{journal}{\emph{Applied Ergonomics}} \bibinfo{volume}{34},
  \bibinfo{number}{6} (\bibinfo{date}{11} \bibinfo{year}{2003}),
  \bibinfo{pages}{499--509}.
\newblock
\showISSN{00036870}
\urldef\tempurl%
\url{https://doi.org/10.1016/j.apergo.2003.07.003}
\showDOI{\tempurl}


\bibitem[B{\"{o}}ckle et~al\mbox{.}(2021)]%
        {Bockle2021CanInterfaces}
\bibfield{author}{\bibinfo{person}{Martin B{\"{o}}ckle}, \bibinfo{person}{Kwaku
  Yeboah-Antwi}, {and} \bibinfo{person}{Iana Kouris}.}
  \bibinfo{year}{2021}\natexlab{}.
\newblock \showarticletitle{{Can You Trust the Black Box? The Effect of
  Personality Traits on Trust in AI-Enabled User Interfaces}}. In
  \bibinfo{booktitle}{\emph{International Conference on Human-Computer
  Interaction}}. \bibinfo{publisher}{Springer}, \bibinfo{pages}{3--20}.
\newblock


\bibitem[Bruckman(2014)]%
        {Bruckman2014ResearchHCI}
\bibfield{author}{\bibinfo{person}{Amy Bruckman}.}
  \bibinfo{year}{2014}\natexlab{}.
\newblock \showarticletitle{{Research ethics and HCI}}.
\newblock In \bibinfo{booktitle}{\emph{Ways of Knowing in HCI}}.
  \bibinfo{publisher}{Springer}, \bibinfo{pages}{449--468}.
\newblock


\bibitem[Brzowski and Nathan-Roberts(2019)]%
        {Brzowski2019TrustReview}
\bibfield{author}{\bibinfo{person}{Matthew Brzowski} {and} \bibinfo{person}{Dan
  Nathan-Roberts}.} \bibinfo{year}{2019}\natexlab{}.
\newblock \showarticletitle{{Trust Measurement in Human–Automation
  Interaction: A Systematic Review}}.
\newblock \bibinfo{journal}{\emph{Proceedings of the Human Factors and
  Ergonomics Society Annual Meeting}} \bibinfo{volume}{63}, \bibinfo{number}{1}
  (\bibinfo{date}{11} \bibinfo{year}{2019}), \bibinfo{pages}{1595--1599}.
\newblock
\showISSN{2169-5067}
\urldef\tempurl%
\url{https://doi.org/10.1177/1071181319631462}
\showDOI{\tempurl}


\bibitem[Burton et~al\mbox{.}(2020)]%
        {Burton2020AMaking}
\bibfield{author}{\bibinfo{person}{Jason~W. Burton},
  \bibinfo{person}{Mari‐Klara Stein}, {and} \bibinfo{person}{Tina~Blegind
  Jensen}.} \bibinfo{year}{2020}\natexlab{}.
\newblock \showarticletitle{{A systematic review of algorithm aversion in
  augmented decision making}}.
\newblock \bibinfo{journal}{\emph{Journal of Behavioral Decision Making}}
  \bibinfo{volume}{33}, \bibinfo{number}{2} (\bibinfo{date}{4}
  \bibinfo{year}{2020}), \bibinfo{pages}{220--239}.
\newblock
\showISSN{0894-3257}
\urldef\tempurl%
\url{https://doi.org/10.1002/bdm.2155}
\showDOI{\tempurl}


\bibitem[Castelfranchi and Falcone(2010)]%
        {Castelfranchi2010_Socio}
\bibfield{author}{\bibinfo{person}{Christiano Castelfranchi} {and}
  \bibinfo{person}{Rino Falcone}.} \bibinfo{year}{2010}\natexlab{}.
\newblock \showarticletitle{{Socio-Cognitive Model of Trust: Basic
  Ingredients}}.
\newblock In \bibinfo{booktitle}{\emph{Trust Theory}}. \bibinfo{publisher}{John
  Wiley {\&} Sons, Ltd}, \bibinfo{address}{Chichester, UK}, Chapter~2,
  \bibinfo{pages}{35--94}.
\newblock
\urldef\tempurl%
\url{https://doi.org/10.1002/9780470519851.ch2}
\showDOI{\tempurl}


\bibitem[Chan-Olmsted(2019)]%
        {Chan-Olmsted2019AIndustry}
\bibfield{author}{\bibinfo{person}{Sylvia~M. Chan-Olmsted}.}
  \bibinfo{year}{2019}\natexlab{}.
\newblock \showarticletitle{{A Review of Artificial Intelligence Adoptions in
  the Media Industry}}.
\newblock \bibinfo{journal}{\emph{International Journal on Media Management}}
  \bibinfo{volume}{21}, \bibinfo{number}{3-4} (\bibinfo{date}{10}
  \bibinfo{year}{2019}), \bibinfo{pages}{193--215}.
\newblock
\showISSN{1424-1277}
\urldef\tempurl%
\url{https://doi.org/10.1080/14241277.2019.1695619}
\showDOI{\tempurl}


\bibitem[Chanley et~al\mbox{.}(2000)]%
        {Chanlet_Consequences_Trust_Government}
\bibfield{author}{\bibinfo{person}{Virginia~A. Chanley},
  \bibinfo{person}{Thomas~J. Rudolph}, {and} \bibinfo{person}{Wendy~M. Rahn}.}
  \bibinfo{year}{2000}\natexlab{}.
\newblock \showarticletitle{{The Origins and Consequences of Public Trust in
  Government}}.
\newblock \bibinfo{journal}{\emph{Public Opinion Quarterly}}
  \bibinfo{volume}{64}, \bibinfo{number}{3} (\bibinfo{year}{2000}),
  \bibinfo{pages}{239--256}.
\newblock
\showISSN{0033362X}
\urldef\tempurl%
\url{https://doi.org/10.1086/317987}
\showDOI{\tempurl}


\bibitem[Cheng et~al\mbox{.}(2022)]%
        {Cheng2022HeartMetaanalysis}
\bibfield{author}{\bibinfo{person}{Ying‐Chih Cheng}, \bibinfo{person}{Min‐I
  Su}, \bibinfo{person}{Cheng‐Wei Liu}, \bibinfo{person}{Yu‐Chen Huang},
  {and} \bibinfo{person}{Wei‐Lieh Huang}.} \bibinfo{year}{2022}\natexlab{}.
\newblock \showarticletitle{{Heart rate variability in patients with anxiety
  disorders: A systematic review and meta‐analysis}}.
\newblock \bibinfo{journal}{\emph{Psychiatry and Clinical Neurosciences}}
  \bibinfo{volume}{76}, \bibinfo{number}{7} (\bibinfo{date}{7}
  \bibinfo{year}{2022}), \bibinfo{pages}{292--302}.
\newblock
\showISSN{1323-1316}
\urldef\tempurl%
\url{https://doi.org/10.1111/pcn.13356}
\showDOI{\tempurl}


\bibitem[Chita-Tegmark et~al\mbox{.}(2021)]%
        {Chita-Tegmark2021CanMeasure}
\bibfield{author}{\bibinfo{person}{Meia Chita-Tegmark},
  \bibinfo{person}{Theresa Law}, \bibinfo{person}{Nicholas Rabb}, {and}
  \bibinfo{person}{Matthias Scheutz}.} \bibinfo{year}{2021}\natexlab{}.
\newblock \showarticletitle{{Can You Trust Your Trust Measure?}}. In
  \bibinfo{booktitle}{\emph{Proceedings of the 2021 ACM/IEEE International
  Conference on Human-Robot Interaction}}. \bibinfo{publisher}{ACM},
  \bibinfo{address}{New York, NY, USA}, \bibinfo{pages}{92--100}.
\newblock
\showISBNx{9781450382892}
\urldef\tempurl%
\url{https://doi.org/10.1145/3434073.3444677}
\showDOI{\tempurl}


\bibitem[Cho et~al\mbox{.}(2006)]%
        {Cho2006OnlineHesitation}
\bibfield{author}{\bibinfo{person}{Chang-Hoan Cho}, \bibinfo{person}{Jaewon
  Kang}, {and} \bibinfo{person}{Hongsik~John Cheon}.}
  \bibinfo{year}{2006}\natexlab{}.
\newblock \showarticletitle{{Online Shopping Hesitation}}.
\newblock \bibinfo{journal}{\emph{CyberPsychology {\&} Behavior}}
  \bibinfo{volume}{9}, \bibinfo{number}{3} (\bibinfo{date}{6}
  \bibinfo{year}{2006}), \bibinfo{pages}{261--274}.
\newblock
\showISSN{1094-9313}
\urldef\tempurl%
\url{https://doi.org/10.1089/cpb.2006.9.261}
\showDOI{\tempurl}


\bibitem[Connor et~al\mbox{.}(2000)]%
        {Connor2000PsychometricScale}
\bibfield{author}{\bibinfo{person}{Kathryn~M Connor}, \bibinfo{person}{Jonathan
  R~T Davidson}, \bibinfo{person}{L~Erik Churchill}, \bibinfo{person}{Andrew
  Sherwood}, \bibinfo{person}{Richard~H Weisler}, {and} \bibinfo{person}{Edna
  Foa}.} \bibinfo{year}{2000}\natexlab{}.
\newblock \showarticletitle{{Psychometric properties of the Social Phobia
  Inventory (SPIN): New self-rating scale}}.
\newblock \bibinfo{journal}{\emph{British Journal of Psychiatry}}
  \bibinfo{volume}{176}, \bibinfo{number}{4} (\bibinfo{year}{2000}),
  \bibinfo{pages}{379--386}.
\newblock
\showISSN{0007-1250}
\urldef\tempurl%
\url{https://doi.org/DOI: 10.1192/bjp.176.4.379}
\showDOI{\tempurl}


\bibitem[Connors et~al\mbox{.}(2015)]%
        {Connors2015Evidence-basedHealth}
\bibfield{author}{\bibinfo{person}{Elizabeth~H. Connors},
  \bibinfo{person}{Prerna Arora}, \bibinfo{person}{Latisha Curtis}, {and}
  \bibinfo{person}{Sharon~H. Stephan}.} \bibinfo{year}{2015}\natexlab{}.
\newblock \showarticletitle{{Evidence-based assessment in school mental
  health}}.
\newblock \bibinfo{journal}{\emph{Cognitive and Behavioral Practice}}
  (\bibinfo{year}{2015}).
\newblock
\showISSN{1878187X}
\urldef\tempurl%
\url{https://doi.org/10.1016/j.cbpra.2014.03.008}
\showDOI{\tempurl}


\bibitem[Coskun and Grabowski(2005)]%
        {Coskun2005ImpactsSystems}
\bibfield{author}{\bibinfo{person}{Erman Coskun} {and} \bibinfo{person}{Martha
  Grabowski}.} \bibinfo{year}{2005}\natexlab{}.
\newblock \showarticletitle{{Impacts of User Interface Complexity on User
  Acceptance and Performance in Safety-Critical Systems}}.
\newblock \bibinfo{journal}{\emph{Journal of Homeland Security and Emergency
  Management}} \bibinfo{volume}{2}, \bibinfo{number}{1} (\bibinfo{date}{1}
  \bibinfo{year}{2005}).
\newblock
\showISSN{1547-7355}
\urldef\tempurl%
\url{https://doi.org/10.2202/1547-7355.1109}
\showDOI{\tempurl}


\bibitem[Damioli et~al\mbox{.}(2021)]%
        {Damioli2021TheProductivity}
\bibfield{author}{\bibinfo{person}{Giacomo Damioli}, \bibinfo{person}{Vincent
  Van~Roy}, {and} \bibinfo{person}{Daniel Vertesy}.}
  \bibinfo{year}{2021}\natexlab{}.
\newblock \showarticletitle{{The impact of artificial intelligence on labor
  productivity}}.
\newblock \bibinfo{journal}{\emph{Eurasian Business Review}}
  \bibinfo{volume}{11}, \bibinfo{number}{1} (\bibinfo{date}{3}
  \bibinfo{year}{2021}), \bibinfo{pages}{1--25}.
\newblock
\showISSN{1309-4297}
\urldef\tempurl%
\url{https://doi.org/10.1007/s40821-020-00172-8}
\showDOI{\tempurl}


\bibitem[Davis(1989)]%
        {Davis1989PerceivedTechnology}
\bibfield{author}{\bibinfo{person}{Fred~D. Davis}.}
  \bibinfo{year}{1989}\natexlab{}.
\newblock \showarticletitle{{Perceived Usefulness, Perceived Ease of Use, and
  User Acceptance of Information Technology}}.
\newblock \bibinfo{journal}{\emph{MIS Quarterly}} \bibinfo{volume}{13},
  \bibinfo{number}{3} (\bibinfo{date}{9} \bibinfo{year}{1989}),
  \bibinfo{pages}{319}.
\newblock
\showISSN{02767783}
\urldef\tempurl%
\url{https://doi.org/10.2307/249008}
\showDOI{\tempurl}


\bibitem[Dechant et~al\mbox{.}(2017)]%
        {Dechant2017}
\bibfield{author}{\bibinfo{person}{M. Dechant}, \bibinfo{person}{S. Trimpl},
  \bibinfo{person}{C. Wolff}, \bibinfo{person}{A. M{\"{u}}hlberger}, {and}
  \bibinfo{person}{Y. Shiban}.} \bibinfo{year}{2017}\natexlab{}.
\newblock \showarticletitle{{Potential of virtual reality as a diagnostic tool
  for social anxiety: A pilot study}}.
\newblock \bibinfo{journal}{\emph{Computers in Human Behavior}}
  \bibinfo{volume}{76} (\bibinfo{year}{2017}).
\newblock
\showISSN{07475632}
\urldef\tempurl%
\url{https://doi.org/10.1016/j.chb.2017.07.005}
\showDOI{\tempurl}


\bibitem[Dechant et~al\mbox{.}(2021)]%
        {Dechant2021TheAnxiety}
\bibfield{author}{\bibinfo{person}{Martin~Johannes Dechant},
  \bibinfo{person}{Julian Frommel}, {and} \bibinfo{person}{Regan~Lee Mandryk}.}
  \bibinfo{year}{2021}\natexlab{}.
\newblock \showarticletitle{{The Development of Explicit and Implicit
  Game-Based Digital Behavioral Markers for the Assessment of Social Anxiety}}.
\newblock \bibinfo{journal}{\emph{Frontiers in Psychology}}
  \bibinfo{volume}{12} (\bibinfo{year}{2021}), \bibinfo{pages}{5566}.
\newblock
\showISSN{1664-1078}
\urldef\tempurl%
\url{https://doi.org/10.3389/fpsyg.2021.760850}
\showDOI{\tempurl}


\bibitem[Dechant et~al\mbox{.}(2022)]%
        {Dechant2022DontTask}
\bibfield{author}{\bibinfo{person}{Martin~Johannes Dechant},
  \bibinfo{person}{Robin Welsch}, \bibinfo{person}{Julian Frommel}, {and}
  \bibinfo{person}{Regan~L Mandryk}.} \bibinfo{year}{2022}\natexlab{}.
\newblock \showarticletitle{{(Don’t) stand by me: How trait psychopathy and
  NPC emotion influence player perceptions, verbal responses, and movement
  behaviours in a gaming task}}. In \bibinfo{booktitle}{\emph{CHI Conference on
  Human Factors in Computing Systems}}. \bibinfo{pages}{1--17}.
\newblock


\bibitem[Dewey(1895)]%
        {Dewey1895TheEmotion.}
\bibfield{author}{\bibinfo{person}{John Dewey}.}
  \bibinfo{year}{1895}\natexlab{}.
\newblock \showarticletitle{{The theory of emotion.}}
\newblock \bibinfo{journal}{\emph{Psychological Review}} \bibinfo{volume}{2},
  \bibinfo{number}{1} (\bibinfo{date}{1} \bibinfo{year}{1895}),
  \bibinfo{pages}{13--32}.
\newblock
\showISSN{1939-1471}
\urldef\tempurl%
\url{https://doi.org/10.1037/h0070927}
\showDOI{\tempurl}


\bibitem[Doney et~al\mbox{.}(1998)]%
        {Doney1998UnderstandingTrust}
\bibfield{author}{\bibinfo{person}{Patricia~M. Doney},
  \bibinfo{person}{Joseph~P. Cannon}, {and} \bibinfo{person}{Michael~R.
  Mullen}.} \bibinfo{year}{1998}\natexlab{}.
\newblock \showarticletitle{{Understanding the Influence of National Culture on
  the Development of Trust}}.
\newblock \bibinfo{journal}{\emph{Academy of Management Review}}
  \bibinfo{volume}{23}, \bibinfo{number}{3} (\bibinfo{date}{7}
  \bibinfo{year}{1998}), \bibinfo{pages}{601--620}.
\newblock
\showISSN{0363-7425}
\urldef\tempurl%
\url{https://doi.org/10.5465/amr.1998.926629}
\showDOI{\tempurl}


\bibitem[Efendi{\'{c}} et~al\mbox{.}(2020)]%
        {Efendic2020SlowPredictions}
\bibfield{author}{\bibinfo{person}{Emir Efendi{\'{c}}},
  \bibinfo{person}{Philippe~P.F.M. Van~de Calseyde}, {and}
  \bibinfo{person}{Anthony~M. Evans}.} \bibinfo{year}{2020}\natexlab{}.
\newblock \showarticletitle{{Slow response times undermine trust in algorithmic
  (but not human) predictions}}.
\newblock \bibinfo{journal}{\emph{Organizational Behavior and Human Decision
  Processes}}  \bibinfo{volume}{157} (\bibinfo{date}{3} \bibinfo{year}{2020}),
  \bibinfo{pages}{103--114}.
\newblock
\showISSN{07495978}
\urldef\tempurl%
\url{https://doi.org/10.1016/j.obhdp.2020.01.008}
\showDOI{\tempurl}


\bibitem[Evans and Krueger(2009)]%
        {Evans2009TheTrust}
\bibfield{author}{\bibinfo{person}{Anthony~M. Evans} {and}
  \bibinfo{person}{Joachim~I. Krueger}.} \bibinfo{year}{2009}\natexlab{}.
\newblock \showarticletitle{{The Psychology (and Economics) of Trust}}.
\newblock \bibinfo{journal}{\emph{Social and Personality Psychology Compass}}
  \bibinfo{volume}{3}, \bibinfo{number}{6} (\bibinfo{date}{12}
  \bibinfo{year}{2009}), \bibinfo{pages}{1003--1017}.
\newblock
\showISSN{17519004}
\urldef\tempurl%
\url{https://doi.org/10.1111/j.1751-9004.2009.00232.x}
\showDOI{\tempurl}


\bibitem[Fessl et~al\mbox{.}(2011)]%
        {Fessl2011MotivationReflection}
\bibfield{author}{\bibinfo{person}{Angela Fessl}, \bibinfo{person}{Verónica
  Rivera-Pelayo}, \bibinfo{person}{Lars M{\"{u}}ller},
  \bibinfo{person}{Viktoria Pammer}, {and} \bibinfo{person}{Stefanie
  Lindstaedt}.} \bibinfo{year}{2011}\natexlab{}.
\newblock \showarticletitle{{Motivation and user acceptance of using
  physiological data to support individual reflection}}. In
  \bibinfo{booktitle}{\emph{2nd International Workshop on Motivation and
  Affective Aspects in Technology Enhanced Learning (MATEL 2011), co-located
  with ECTEL}}.
\newblock


\bibitem[Floridi and Taddeo(2016)]%
        {Floridi2016}
\bibfield{author}{\bibinfo{person}{Luciano Floridi} {and}
  \bibinfo{person}{Mariarosaria Taddeo}.} \bibinfo{year}{2016}\natexlab{}.
\newblock \showarticletitle{{What is data ethics?}}
\newblock \bibinfo{journal}{\emph{Philosophical Transactions of the Royal
  Society A: Mathematical, Physical and Engineering Sciences}}
  \bibinfo{volume}{374}, \bibinfo{number}{2083} (\bibinfo{date}{12}
  \bibinfo{year}{2016}), \bibinfo{pages}{20160360}.
\newblock
\showISSN{1364-503X}
\urldef\tempurl%
\url{https://doi.org/10.1098/rsta.2016.0360}
\showDOI{\tempurl}


\bibitem[Freeman et~al\mbox{.}(2017)]%
        {Freeman2017VirtualDisorders}
\bibfield{author}{\bibinfo{person}{D. Freeman}, \bibinfo{person}{S. Reeve},
  \bibinfo{person}{A. Robinson}, \bibinfo{person}{A. Ehlers},
  \bibinfo{person}{D. Clark}, \bibinfo{person}{B. Spanlang}, {and}
  \bibinfo{person}{M. Slater}.} \bibinfo{year}{2017}\natexlab{}.
\newblock \bibinfo{title}{{Virtual reality in the assessment, understanding,
  and treatment of mental health disorders}}.
\newblock
\newblock
\showISSN{14698978}
\urldef\tempurl%
\url{https://doi.org/10.1017/S003329171700040X}
\showDOI{\tempurl}


\bibitem[Freeman(2018)]%
        {Freeman2018DoingHand}
\bibfield{author}{\bibinfo{person}{Jonathan~B. Freeman}.}
  \bibinfo{year}{2018}\natexlab{}.
\newblock \showarticletitle{{Doing Psychological Science by Hand}}.
\newblock \bibinfo{journal}{\emph{Current Directions in Psychological Science}}
  \bibinfo{volume}{27}, \bibinfo{number}{5} (\bibinfo{date}{10}
  \bibinfo{year}{2018}), \bibinfo{pages}{315--323}.
\newblock
\showISSN{0963-7214}
\urldef\tempurl%
\url{https://doi.org/10.1177/0963721417746793}
\showDOI{\tempurl}


\bibitem[Gilbert(1998)]%
        {Gilbert1998TowardsTrust}
\bibfield{author}{\bibinfo{person}{Tony Gilbert}.}
  \bibinfo{year}{1998}\natexlab{}.
\newblock \showarticletitle{{Towards a politics of trust}}.
\newblock \bibinfo{journal}{\emph{Journal of Advanced Nursing}}
  \bibinfo{volume}{27}, \bibinfo{number}{5} (\bibinfo{date}{5}
  \bibinfo{year}{1998}), \bibinfo{pages}{1010--1016}.
\newblock
\showISSN{0309-2402}
\urldef\tempurl%
\url{https://doi.org/10.1046/j.1365-2648.1998.00578.x}
\showDOI{\tempurl}


\bibitem[Gilbert(2005)]%
        {Gilbert2005ImpersonalDynamics}
\bibfield{author}{\bibinfo{person}{Tony~P. Gilbert}.}
  \bibinfo{year}{2005}\natexlab{}.
\newblock \showarticletitle{{Impersonal trust and professional authority:
  exploring the dynamics}}.
\newblock \bibinfo{journal}{\emph{Journal of Advanced Nursing}}
  \bibinfo{volume}{49}, \bibinfo{number}{6} (\bibinfo{date}{3}
  \bibinfo{year}{2005}), \bibinfo{pages}{568--577}.
\newblock
\showISSN{0309-2402}
\urldef\tempurl%
\url{https://doi.org/10.1111/j.1365-2648.2004.03332.x}
\showDOI{\tempurl}


\bibitem[Glikson and Woolley(2020)]%
        {Glikson2020HumanResearch}
\bibfield{author}{\bibinfo{person}{Ella Glikson} {and}
  \bibinfo{person}{Anita~Williams Woolley}.} \bibinfo{year}{2020}\natexlab{}.
\newblock \showarticletitle{{Human Trust in Artificial Intelligence: Review of
  Empirical Research}}.
\newblock \bibinfo{journal}{\emph{Academy of Management Annals}}
  \bibinfo{volume}{14}, \bibinfo{number}{2} (\bibinfo{date}{7}
  \bibinfo{year}{2020}), \bibinfo{pages}{627--660}.
\newblock
\showISSN{1941-6520}
\urldef\tempurl%
\url{https://doi.org/10.5465/annals.2018.0057}
\showDOI{\tempurl}


\bibitem[Gunning et~al\mbox{.}(2019)]%
        {Gunning2019XAIExplainableIntelligence}
\bibfield{author}{\bibinfo{person}{David Gunning}, \bibinfo{person}{Mark
  Stefik}, \bibinfo{person}{Jaesik Choi}, \bibinfo{person}{Timothy Miller},
  \bibinfo{person}{Simone Stumpf}, {and} \bibinfo{person}{Guang-Zhong Yang}.}
  \bibinfo{year}{2019}\natexlab{}.
\newblock \showarticletitle{{XAI—Explainable artificial intelligence}}.
\newblock \bibinfo{journal}{\emph{Science Robotics}} \bibinfo{volume}{4},
  \bibinfo{number}{37} (\bibinfo{date}{12} \bibinfo{year}{2019}).
\newblock
\showISSN{2470-9476}
\urldef\tempurl%
\url{https://doi.org/10.1126/scirobotics.aay7120}
\showDOI{\tempurl}


\bibitem[Gupta et~al\mbox{.}(2020)]%
        {Gupta2020MeasuringReality}
\bibfield{author}{\bibinfo{person}{Kunal Gupta}, \bibinfo{person}{Ryo Hajika},
  \bibinfo{person}{Yun~Suen Pai}, \bibinfo{person}{Andreas Duenser},
  \bibinfo{person}{Martin Lochner}, {and} \bibinfo{person}{Mark Billinghurst}.}
  \bibinfo{year}{2020}\natexlab{}.
\newblock \showarticletitle{{Measuring Human Trust in a Virtual Assistant using
  Physiological Sensing in Virtual Reality}}. In \bibinfo{booktitle}{\emph{2020
  IEEE Conference on Virtual Reality and 3D User Interfaces (VR)}}.
  \bibinfo{publisher}{IEEE}, \bibinfo{pages}{756--765}.
\newblock
\showISBNx{978-1-7281-5608-8}
\urldef\tempurl%
\url{https://doi.org/10.1109/VR46266.2020.00099}
\showDOI{\tempurl}


\bibitem[Harrison~McKnight and Chervany(2001)]%
        {HarrisonMcKnight2001TrustTime}
\bibfield{author}{\bibinfo{person}{D. Harrison~McKnight} {and}
  \bibinfo{person}{Norman~L. Chervany}.} \bibinfo{year}{2001}\natexlab{}.
\newblock \showarticletitle{{Trust and Distrust Definitions: One Bite at a
  Time}}.
\newblock \bibinfo{pages}{27--54}.
\newblock
\urldef\tempurl%
\url{https://doi.org/10.1007/3-540-45547-7{\_}3}
\showDOI{\tempurl}


\bibitem[Hastie et~al\mbox{.}(2017)]%
        {Hastie2017TrustInterfaces}
\bibfield{author}{\bibinfo{person}{Helen Hastie}, \bibinfo{person}{Xingkun
  Liu}, {and} \bibinfo{person}{Pedro Patron}.} \bibinfo{year}{2017}\natexlab{}.
\newblock \showarticletitle{{Trust triggers for multimodal command and control
  interfaces}}. In \bibinfo{booktitle}{\emph{Proceedings of the 19th ACM
  International Conference on Multimodal Interaction}}.
  \bibinfo{publisher}{ACM}, \bibinfo{address}{New York, NY, USA},
  \bibinfo{pages}{261--268}.
\newblock
\showISBNx{9781450355438}
\urldef\tempurl%
\url{https://doi.org/10.1145/3136755.3136764}
\showDOI{\tempurl}


\bibitem[Hern{\'{a}}ndez-Ortega(2011)]%
        {Hernandez-Ortega2011TheConsequences}
\bibfield{author}{\bibinfo{person}{Blanca Hern{\'{a}}ndez-Ortega}.}
  \bibinfo{year}{2011}\natexlab{}.
\newblock \showarticletitle{{The role of post-use trust in the acceptance of a
  technology: Drivers and consequences}}.
\newblock \bibinfo{journal}{\emph{Technovation}} \bibinfo{volume}{31},
  \bibinfo{number}{10-11} (\bibinfo{date}{10} \bibinfo{year}{2011}),
  \bibinfo{pages}{523--538}.
\newblock
\showISSN{01664972}
\urldef\tempurl%
\url{https://doi.org/10.1016/j.technovation.2011.07.001}
\showDOI{\tempurl}


\bibitem[Hosmer(1995)]%
        {Hosmer1995Trust:Ethics}
\bibfield{author}{\bibinfo{person}{Larue~Tone Hosmer}.}
  \bibinfo{year}{1995}\natexlab{}.
\newblock \showarticletitle{{Trust: The Connecting Link between Organizational
  Theory and Philosophical Ethics}}.
\newblock \bibinfo{journal}{\emph{The Academy of Management Review}}
  \bibinfo{volume}{20}, \bibinfo{number}{2} (\bibinfo{date}{4}
  \bibinfo{year}{1995}), \bibinfo{pages}{379}.
\newblock
\showISSN{03637425}
\urldef\tempurl%
\url{https://doi.org/10.2307/258851}
\showDOI{\tempurl}


\bibitem[Hosp et~al\mbox{.}(2021b)]%
        {Hosp2021StatesSurgery}
\bibfield{author}{\bibinfo{person}{Benedikt Hosp}, \bibinfo{person}{Myat~Su
  Yin}, \bibinfo{person}{Peter Haddawy}, \bibinfo{person}{Ratthaphum
  Watcharopas}, \bibinfo{person}{Paphon Sa-Ngasoongsong}, {and}
  \bibinfo{person}{Enkelejda Kasneci}.} \bibinfo{year}{2021}\natexlab{b}.
\newblock \showarticletitle{{States of Confusion: Eye and Head Tracking Reveal
  Surgeons’ Confusion during Arthroscopic Surgery}}. In
  \bibinfo{booktitle}{\emph{Proceedings of the 2021 International Conference on
  Multimodal Interaction}}. \bibinfo{publisher}{ACM}, \bibinfo{address}{New
  York, NY, USA}, \bibinfo{pages}{753--757}.
\newblock
\showISBNx{9781450384810}
\urldef\tempurl%
\url{https://doi.org/10.1145/3462244.3479953}
\showDOI{\tempurl}


\bibitem[Hosp et~al\mbox{.}(2021a)]%
        {Hosp2021SoccerMovements}
\bibfield{author}{\bibinfo{person}{Benedikt~W. Hosp}, \bibinfo{person}{Florian
  Schultz}, \bibinfo{person}{Oliver H{\"{o}}ner}, {and}
  \bibinfo{person}{Enkelejda Kasneci}.} \bibinfo{year}{2021}\natexlab{a}.
\newblock \showarticletitle{{Soccer goalkeeper expertise identification based
  on eye movements}}.
\newblock \bibinfo{journal}{\emph{PLOS ONE}} \bibinfo{volume}{16},
  \bibinfo{number}{5} (\bibinfo{date}{5} \bibinfo{year}{2021}),
  \bibinfo{pages}{e0251070}.
\newblock
\showISSN{1932-6203}
\urldef\tempurl%
\url{https://doi.org/10.1371/journal.pone.0251070}
\showDOI{\tempurl}


\bibitem[Hu et~al\mbox{.}(2016)]%
        {Hu2016Real-TimeFoundation.}
\bibfield{author}{\bibinfo{person}{Wan-Lin Hu}, \bibinfo{person}{Kumar Akash},
  \bibinfo{person}{Neera Jain}, {and} \bibinfo{person}{Tahira Reid}.}
  \bibinfo{year}{2016}\natexlab{}.
\newblock \showarticletitle{{Real-Time Sensing of Trust in Human-Machine
  Interactions**This material is based upon work supported by the National
  Science Foundation under Award No. 1548616. Any opinions, findings, and
  conclusions or recommendations expressed in this material are those of the
  author(s) and do not necessarily reflect the views of the National Science
  Foundation.}}
\newblock \bibinfo{journal}{\emph{IFAC-PapersOnLine}} \bibinfo{volume}{49},
  \bibinfo{number}{32} (\bibinfo{year}{2016}), \bibinfo{pages}{48--53}.
\newblock
\showISSN{24058963}
\urldef\tempurl%
\url{https://doi.org/10.1016/j.ifacol.2016.12.188}
\showDOI{\tempurl}


\bibitem[Hunsley and Mash(2007)]%
        {Hunsley2007Evidence-BasedAssessment}
\bibfield{author}{\bibinfo{person}{John Hunsley} {and} \bibinfo{person}{Eric~J.
  Mash}.} \bibinfo{year}{2007}\natexlab{}.
\newblock \showarticletitle{{Evidence-Based Assessment}}.
\newblock \bibinfo{journal}{\emph{Annual Review of Clinical Psychology}}
  (\bibinfo{year}{2007}).
\newblock
\showISBNx{0824339037}
\showISSN{1548-5943}
\urldef\tempurl%
\url{https://doi.org/10.1146/annurev.clinpsy.3.022806.091419}
\showDOI{\tempurl}


\bibitem[Jarvenpaa et~al\mbox{.}(2000)]%
        {Jarvenpaa2000ConsumerStore}
\bibfield{author}{\bibinfo{person}{Sirkka~L. Jarvenpaa}, \bibinfo{person}{Noam
  Tractinsky}, {and} \bibinfo{person}{Michael Vitale}.}
  \bibinfo{year}{2000}\natexlab{}.
\newblock \showarticletitle{{Consumer trust in an Internet store}}.
\newblock \bibinfo{journal}{\emph{INFORMATION TECHNOLOGY AND MANAGEMENT}}
  \bibinfo{volume}{1}, \bibinfo{number}{1/2} (\bibinfo{year}{2000}),
  \bibinfo{pages}{45--71}.
\newblock
\showISSN{1385951X}
\urldef\tempurl%
\url{https://doi.org/10.1023/A:1019104520776}
\showDOI{\tempurl}


\bibitem[Jay and Janschewitz(2008)]%
        {Jay2008TheSwearing}
\bibfield{author}{\bibinfo{person}{Timothy Jay} {and} \bibinfo{person}{Kristin
  Janschewitz}.} \bibinfo{year}{2008}\natexlab{}.
\newblock \showarticletitle{{The pragmatics of swearing}}.
\newblock \bibinfo{journal}{\emph{Journal of Politeness Research. Language,
  Behaviour, Culture}} \bibinfo{volume}{4}, \bibinfo{number}{2}
  (\bibinfo{date}{1} \bibinfo{year}{2008}).
\newblock
\showISSN{1612-5681}
\urldef\tempurl%
\url{https://doi.org/10.1515/JPLR.2008.013}
\showDOI{\tempurl}


\bibitem[Johannes~Dechant et~al\mbox{.}(2021)]%
        {JohannesDechant2021AssessingTask}
\bibfield{author}{\bibinfo{person}{Martin Johannes~Dechant},
  \bibinfo{person}{Julian Frommel}, {and} \bibinfo{person}{Regan Mandryk}.}
  \bibinfo{year}{2021}\natexlab{}.
\newblock \showarticletitle{{Assessing social anxiety through digital
  biomarkers embedded in a gaming task}}. In
  \bibinfo{booktitle}{\emph{Proceedings of the 2021 CHI Conference on Human
  Factors in Computing Systems}}. \bibinfo{pages}{1--15}.
\newblock


\bibitem[Johanson(2020)]%
        {Johanson2020}
\bibfield{author}{\bibinfo{person}{Colby Johanson}.}
  \bibinfo{year}{2020}\natexlab{}.
\newblock \bibinfo{title}{{colbyj/bride-of-frankensystem 1.1}}.
\newblock
\newblock
\urldef\tempurl%
\url{https://doi.org/10.5281/ZENODO.3738761}
\showDOI{\tempurl}


\bibitem[Johnson and Verdicchio(2017)]%
        {Johnson2017AIAnxiety}
\bibfield{author}{\bibinfo{person}{Deborah~G. Johnson} {and}
  \bibinfo{person}{Mario Verdicchio}.} \bibinfo{year}{2017}\natexlab{}.
\newblock \showarticletitle{{AI Anxiety}}.
\newblock \bibinfo{journal}{\emph{Journal of the Association for Information
  Science and Technology}} \bibinfo{volume}{68}, \bibinfo{number}{9}
  (\bibinfo{date}{9} \bibinfo{year}{2017}), \bibinfo{pages}{2267--2270}.
\newblock
\showISSN{23301635}
\urldef\tempurl%
\url{https://doi.org/10.1002/asi.23867}
\showDOI{\tempurl}


\bibitem[Kaur et~al\mbox{.}(2023)]%
        {Kaur2023TrustworthyReview}
\bibfield{author}{\bibinfo{person}{Davinder Kaur}, \bibinfo{person}{Suleyman
  Uslu}, \bibinfo{person}{Kaley~J. Rittichier}, {and} \bibinfo{person}{Arjan
  Durresi}.} \bibinfo{year}{2023}\natexlab{}.
\newblock \showarticletitle{{Trustworthy Artificial Intelligence: A Review}}.
\newblock \bibinfo{journal}{\emph{Comput. Surveys}} \bibinfo{volume}{55},
  \bibinfo{number}{2} (\bibinfo{date}{3} \bibinfo{year}{2023}),
  \bibinfo{pages}{1--38}.
\newblock
\showISSN{0360-0300}
\urldef\tempurl%
\url{https://doi.org/10.1145/3491209}
\showDOI{\tempurl}


\bibitem[Kieslich and Henninger(2017)]%
        {Kieslich2017Mousetrap:Package}
\bibfield{author}{\bibinfo{person}{Pascal~J. Kieslich} {and}
  \bibinfo{person}{Felix Henninger}.} \bibinfo{year}{2017}\natexlab{}.
\newblock \showarticletitle{{Mousetrap: An integrated, open-source
  mouse-tracking package}}.
\newblock \bibinfo{journal}{\emph{Behavior Research Methods}}
  \bibinfo{volume}{49}, \bibinfo{number}{5} (\bibinfo{date}{10}
  \bibinfo{year}{2017}), \bibinfo{pages}{1652--1667}.
\newblock
\showISSN{1554-3528}
\urldef\tempurl%
\url{https://doi.org/10.3758/s13428-017-0900-z}
\showDOI{\tempurl}


\bibitem[Knight(1921)]%
        {Knight1921RiskProfit}
\bibfield{author}{\bibinfo{person}{Frank~Hyneman Knight}.}
  \bibinfo{year}{1921}\natexlab{}.
\newblock \bibinfo{booktitle}{\emph{{Risk, uncertainty and profit}}}.
  Vol.~\bibinfo{volume}{31}.
\newblock \bibinfo{publisher}{Houghton Mifflin}.
\newblock


\bibitem[Kohn et~al\mbox{.}(2021)]%
        {Kohn2021MeasurementGuide}
\bibfield{author}{\bibinfo{person}{Spencer~C. Kohn}, \bibinfo{person}{Ewart~J.
  de Visser}, \bibinfo{person}{Eva Wiese}, \bibinfo{person}{Yi~Ching Lee},
  {and} \bibinfo{person}{Tyler~H. Shaw}.} \bibinfo{year}{2021}\natexlab{}.
\newblock \bibinfo{title}{{Measurement of Trust in Automation: A Narrative
  Review and Reference Guide}}.
\newblock
\newblock
\showISSN{16641078}
\urldef\tempurl%
\url{https://doi.org/10.3389/fpsyg.2021.604977}
\showDOI{\tempurl}


\bibitem[Lachman and Joffe(2021)]%
        {Lachman2021ApplicationsEntertainment}
\bibfield{author}{\bibinfo{person}{Richard Lachman} {and}
  \bibinfo{person}{Michael Joffe}.} \bibinfo{year}{2021}\natexlab{}.
\newblock \showarticletitle{{Applications of Artificial Intelligence in Media
  and Entertainment}}.
\newblock \bibinfo{pages}{201--220}.
\newblock
\urldef\tempurl%
\url{https://doi.org/10.4018/978-1-7998-3499-1.ch012}
\showDOI{\tempurl}


\bibitem[Lee and See(2004)]%
        {Lee2004TrustReliance}
\bibfield{author}{\bibinfo{person}{John~D. Lee} {and}
  \bibinfo{person}{Katrina~A. See}.} \bibinfo{year}{2004}\natexlab{}.
\newblock \showarticletitle{{Trust in Automation: Designing for Appropriate
  Reliance}}.
\newblock \bibinfo{journal}{\emph{Human Factors: The Journal of the Human
  Factors and Ergonomics Society}} \bibinfo{volume}{46}, \bibinfo{number}{1}
  (\bibinfo{year}{2004}), \bibinfo{pages}{50--80}.
\newblock
\showISSN{1547-8181}
\urldef\tempurl%
\url{https://doi.org/10.1518/hfes.46.1.50.30392}
\showDOI{\tempurl}


\bibitem[Leichtenstern et~al\mbox{.}(2011)]%
        {Leichtenstern2011PhysiologicalSituations}
\bibfield{author}{\bibinfo{person}{Karin Leichtenstern},
  \bibinfo{person}{Nikolaus Bee}, \bibinfo{person}{Elisabeth Andr{\'{e}}},
  \bibinfo{person}{Ulrich Berkm{\"{u}}ller}, {and} \bibinfo{person}{Johannes
  Wagner}.} \bibinfo{year}{2011}\natexlab{}.
\newblock \showarticletitle{{Physiological measurement of trust-related
  behavior in trust-neutral and trust-critical situations}}. In
  \bibinfo{booktitle}{\emph{IFIP Advances in Information and Communication
  Technology}}, Vol.~\bibinfo{volume}{358 AICT}.
\newblock
\showISSN{18684238}
\urldef\tempurl%
\url{https://doi.org/10.1007/978-3-642-22200-9{\_}14}
\showDOI{\tempurl}


\bibitem[Loo(2002)]%
        {Loo2002AScales}
\bibfield{author}{\bibinfo{person}{Robert Loo}.}
  \bibinfo{year}{2002}\natexlab{}.
\newblock \showarticletitle{{A caveat on using single‐item versus
  multiple‐item scales}}.
\newblock \bibinfo{journal}{\emph{Journal of Managerial Psychology}}
  \bibinfo{volume}{17}, \bibinfo{number}{1} (\bibinfo{date}{2}
  \bibinfo{year}{2002}), \bibinfo{pages}{68--75}.
\newblock
\showISSN{0268-3946}
\urldef\tempurl%
\url{https://doi.org/10.1108/02683940210415933}
\showDOI{\tempurl}


\bibitem[Lu and Sarter(2019)]%
        {Lu2019EyeReliability}
\bibfield{author}{\bibinfo{person}{Y Lu} {and} \bibinfo{person}{N Sarter}.}
  \bibinfo{year}{2019}\natexlab{}.
\newblock \showarticletitle{{Eye Tracking: A Process-Oriented Method for
  Inferring Trust in Automation as a Function of Priming and System
  Reliability}}.
\newblock \bibinfo{journal}{\emph{IEEE Transactions on Human-Machine Systems}}
  \bibinfo{volume}{49}, \bibinfo{number}{6} (\bibinfo{year}{2019}),
  \bibinfo{pages}{560--568}.
\newblock
\showISSN{2168-2305}
\urldef\tempurl%
\url{https://doi.org/10.1109/THMS.2019.2930980}
\showDOI{\tempurl}


\bibitem[Luhmann and Gambetta(1988)]%
        {Luhmann1988Trust:Relations}
\bibfield{author}{\bibinfo{person}{Niklas Luhmann} {and} \bibinfo{person}{Diego
  Gambetta}.} \bibinfo{year}{1988}\natexlab{}.
\newblock \showarticletitle{{Trust: making and breaking cooperative
  relations}}.
\newblock \bibinfo{journal}{\emph{Familiarity, confidence, trust: problems and
  alternatives. New York: Basil Blackwell}} (\bibinfo{year}{1988}),
  \bibinfo{pages}{94--107}.
\newblock


\bibitem[Mandryk and Birk(2019)]%
        {Mandryk2019}
\bibfield{author}{\bibinfo{person}{Regan~Lee Mandryk} {and}
  \bibinfo{person}{Max~Valentin Birk}.} \bibinfo{year}{2019}\natexlab{}.
\newblock \bibinfo{title}{{The potential of game-based digital biomarkers for
  modeling mental health}}.
\newblock
\newblock
\showISSN{14388871}
\urldef\tempurl%
\url{https://doi.org/10.2196/13485}
\showDOI{\tempurl}


\bibitem[Mash and Hunsley(2005)]%
        {Mash2005Evidence-basedChallenges}
\bibfield{author}{\bibinfo{person}{Eric~J. Mash} {and} \bibinfo{person}{John
  Hunsley}.} \bibinfo{year}{2005}\natexlab{}.
\newblock \showarticletitle{{Evidence-based assessment of child and adolescent
  disorders: Issues and challenges}}.
\newblock \bibinfo{journal}{\emph{Journal of Clinical Child and Adolescent
  Psychology}} (\bibinfo{year}{2005}).
\newblock
\showISSN{15374416}
\urldef\tempurl%
\url{https://doi.org/10.1207/s15374424jccp3403{\_}1}
\showDOI{\tempurl}


\bibitem[Merrill and Cheshire(2017)]%
        {Merrill2017TrustInteractions}
\bibfield{author}{\bibinfo{person}{Nick Merrill} {and} \bibinfo{person}{Coye
  Cheshire}.} \bibinfo{year}{2017}\natexlab{}.
\newblock \showarticletitle{{Trust Your Heart: Assessing Cooperation and Trust
  with Biosignals in Computer-Mediated Interactions}}. In
  \bibinfo{booktitle}{\emph{Proceedings of the 2017 ACM Conference on Computer
  Supported Cooperative Work and Social Computing}}. \bibinfo{publisher}{ACM},
  \bibinfo{address}{New York, NY, USA}, \bibinfo{pages}{2--12}.
\newblock
\showISBNx{9781450343350}
\urldef\tempurl%
\url{https://doi.org/10.1145/2998181.2998286}
\showDOI{\tempurl}


\bibitem[Merritt et~al\mbox{.}(2013)]%
        {Merritt2013ISystem}
\bibfield{author}{\bibinfo{person}{Stephanie~M. Merritt},
  \bibinfo{person}{Heather Heimbaugh}, \bibinfo{person}{Jennifer LaChapell},
  {and} \bibinfo{person}{Deborah Lee}.} \bibinfo{year}{2013}\natexlab{}.
\newblock \showarticletitle{{I Trust It, but I Don’t Know Why: Effects of
  Implicit Attitudes Toward Automation on Trust in an Automated System}}.
\newblock \bibinfo{journal}{\emph{Human Factors: The Journal of the Human
  Factors and Ergonomics Society}} \bibinfo{volume}{55}, \bibinfo{number}{3}
  (\bibinfo{date}{6} \bibinfo{year}{2013}), \bibinfo{pages}{520--534}.
\newblock
\showISSN{0018-7208}
\urldef\tempurl%
\url{https://doi.org/10.1177/0018720812465081}
\showDOI{\tempurl}


\bibitem[Merritt and Ilgen(2008)]%
        {Merritt2008NotInteractions}
\bibfield{author}{\bibinfo{person}{Stephanie~M. Merritt} {and}
  \bibinfo{person}{Daniel~R. Ilgen}.} \bibinfo{year}{2008}\natexlab{}.
\newblock \showarticletitle{{Not All Trust Is Created Equal: Dispositional and
  History-Based Trust in Human-Automation Interactions}}.
\newblock \bibinfo{journal}{\emph{Human Factors: The Journal of the Human
  Factors and Ergonomics Society}} \bibinfo{volume}{50}, \bibinfo{number}{2}
  (\bibinfo{date}{4} \bibinfo{year}{2008}), \bibinfo{pages}{194--210}.
\newblock
\showISSN{0018-7208}
\urldef\tempurl%
\url{https://doi.org/10.1518/001872008X288574}
\showDOI{\tempurl}


\bibitem[Mittelstadt et~al\mbox{.}(2016)]%
        {Mittelstadt2016}
\bibfield{author}{\bibinfo{person}{Brent~Daniel Mittelstadt},
  \bibinfo{person}{Patrick Allo}, \bibinfo{person}{Mariarosaria Taddeo},
  \bibinfo{person}{Sandra Wachter}, {and} \bibinfo{person}{Luciano Floridi}.}
  \bibinfo{year}{2016}\natexlab{}.
\newblock \showarticletitle{{The ethics of algorithms: Mapping the debate}}.
\newblock \bibinfo{journal}{\emph{Big Data {\&} Society}} \bibinfo{volume}{3},
  \bibinfo{number}{2} (\bibinfo{year}{2016}),
  \bibinfo{pages}{2053951716679679}.
\newblock
\showISSN{2053-9517}


\bibitem[Morris et~al\mbox{.}(2017)]%
        {Morris2017ElectrodermalUse}
\bibfield{author}{\bibinfo{person}{Drew~M. Morris}, \bibinfo{person}{Jason~M.
  Erno}, {and} \bibinfo{person}{June~J. Pilcher}.}
  \bibinfo{year}{2017}\natexlab{}.
\newblock \showarticletitle{{Electrodermal Response and Automation Trust during
  Simulated Self-Driving Car Use}}.
\newblock \bibinfo{journal}{\emph{Proceedings of the Human Factors and
  Ergonomics Society Annual Meeting}} \bibinfo{volume}{61}, \bibinfo{number}{1}
  (\bibinfo{date}{9} \bibinfo{year}{2017}), \bibinfo{pages}{1759--1762}.
\newblock
\showISSN{2169-5067}
\urldef\tempurl%
\url{https://doi.org/10.1177/1541931213601921}
\showDOI{\tempurl}


\bibitem[Narbona et~al\mbox{.}(2020)]%
        {Narbona_FragileTenetsOfTrust}
\bibfield{author}{\bibinfo{person}{Juan Narbona}, \bibinfo{person}{Jordi
  Pujol}, {and} \bibinfo{person}{Anne Gregory}.}
  \bibinfo{year}{2020}\natexlab{}.
\newblock \showarticletitle{{The fragile tenets of trust}}.
\newblock \bibinfo{journal}{\emph{Church, Communication and Culture}}
  \bibinfo{volume}{5}, \bibinfo{number}{3} (\bibinfo{date}{9}
  \bibinfo{year}{2020}), \bibinfo{pages}{293--297}.
\newblock
\showISSN{2375-3234}
\urldef\tempurl%
\url{https://doi.org/10.1080/23753234.2020.1825975}
\showDOI{\tempurl}


\bibitem[O’Cass and Fenech(2003)]%
        {OCass2003WebBehaviour}
\bibfield{author}{\bibinfo{person}{Aron O’Cass} {and} \bibinfo{person}{Tino
  Fenech}.} \bibinfo{year}{2003}\natexlab{}.
\newblock \showarticletitle{{Web retailing adoption: exploring the nature of
  internet users Web retailing behaviour}}.
\newblock \bibinfo{journal}{\emph{Journal of Retailing and Consumer Services}}
  \bibinfo{volume}{10}, \bibinfo{number}{2} (\bibinfo{date}{3}
  \bibinfo{year}{2003}), \bibinfo{pages}{81--94}.
\newblock
\showISSN{09696989}
\urldef\tempurl%
\url{https://doi.org/10.1016/S0969-6989(02)00004-8}
\showDOI{\tempurl}


\bibitem[Papenmeier et~al\mbox{.}(2022)]%
        {Papenmeier2022ItsAI}
\bibfield{author}{\bibinfo{person}{Andrea Papenmeier}, \bibinfo{person}{Dagmar
  Kern}, \bibinfo{person}{Gwenn Englebienne}, {and} \bibinfo{person}{Christin
  Seifert}.} \bibinfo{year}{2022}\natexlab{}.
\newblock \showarticletitle{{It’s Complicated: The Relationship between User
  Trust, Model Accuracy and Explanations in AI}}.
\newblock \bibinfo{journal}{\emph{ACM Transactions on Computer-Human
  Interaction}} \bibinfo{volume}{29}, \bibinfo{number}{4} (\bibinfo{date}{8}
  \bibinfo{year}{2022}), \bibinfo{pages}{1--33}.
\newblock
\showISSN{1073-0516}
\urldef\tempurl%
\url{https://doi.org/10.1145/3495013}
\showDOI{\tempurl}


\bibitem[programme on~chemical safety and mondiale~della sanit{\`{a}}(2001)]%
        {international2001biomarkers}
\bibfield{author}{\bibinfo{person}{International programme on~chemical safety}
  {and} \bibinfo{person}{Organizzazione mondiale~della sanit{\`{a}}}.}
  \bibinfo{year}{2001}\natexlab{}.
\newblock \bibinfo{booktitle}{\emph{{Biomarkers in Risk Assessment: Validity
  and Validation}}}.
\newblock \bibinfo{publisher}{World health organization}.
\newblock


\bibitem[{Prolific}(2014)]%
        {Prolific2014Prolific}
\bibfield{author}{\bibinfo{person}{{Prolific}}.}
  \bibinfo{year}{2014}\natexlab{}.
\newblock \bibinfo{title}{{Prolific}}.
\newblock
\newblock
\urldef\tempurl%
\url{https://prolific.co/}
\showURL{%
\tempurl}


\bibitem[Pu and Chen(2007)]%
        {Pu2007Trust-inspiringSystems}
\bibfield{author}{\bibinfo{person}{Pearl Pu} {and} \bibinfo{person}{Li Chen}.}
  \bibinfo{year}{2007}\natexlab{}.
\newblock \showarticletitle{{Trust-inspiring explanation interfaces for
  recommender systems}}.
\newblock \bibinfo{journal}{\emph{Knowledge-Based Systems}}
  \bibinfo{volume}{20}, \bibinfo{number}{6} (\bibinfo{date}{8}
  \bibinfo{year}{2007}), \bibinfo{pages}{542--556}.
\newblock
\showISSN{09507051}
\urldef\tempurl%
\url{https://doi.org/10.1016/j.knosys.2007.04.004}
\showDOI{\tempurl}


\bibitem[Rajaonah et~al\mbox{.}(2006)]%
        {Rajaonah2006TrustControl}
\bibfield{author}{\bibinfo{person}{Bako Rajaonah}, \bibinfo{person}{Franoise
  Anceaux}, \bibinfo{person}{Nicolas Tricot}, {and}
  \bibinfo{person}{Marie-Pierre Pacaux-Lemoine}.}
  \bibinfo{year}{2006}\natexlab{}.
\newblock \showarticletitle{{Trust, cognitive control, and control: the case of
  drivers using an Auto-Adaptive Cruise Control}}. In
  \bibinfo{booktitle}{\emph{Proceedings of the 13th Eurpoean conference on
  Cognitive ergonomics trust and control in complex socio-technical systems -
  ECCE '06}}. \bibinfo{publisher}{ACM Press}, \bibinfo{address}{New York, New
  York, USA}, \bibinfo{pages}{17}.
\newblock
\showISBNx{9783906509235}
\urldef\tempurl%
\url{https://doi.org/10.1145/1274892.1274896}
\showDOI{\tempurl}


\bibitem[Rheem et~al\mbox{.}(2018)]%
        {Rheem2018UseLoad}
\bibfield{author}{\bibinfo{person}{Hansol Rheem}, \bibinfo{person}{Vipin
  Verma}, {and} \bibinfo{person}{D.~Vaughn Becker}.}
  \bibinfo{year}{2018}\natexlab{}.
\newblock \showarticletitle{{Use of Mouse-tracking Method to Measure Cognitive
  Load}}.
\newblock \bibinfo{journal}{\emph{Proceedings of the Human Factors and
  Ergonomics Society Annual Meeting}} \bibinfo{volume}{62}, \bibinfo{number}{1}
  (\bibinfo{date}{9} \bibinfo{year}{2018}), \bibinfo{pages}{1982--1986}.
\newblock
\showISSN{2169-5067}
\urldef\tempurl%
\url{https://doi.org/10.1177/1541931218621449}
\showDOI{\tempurl}


\bibitem[Riedl and Javor(2012)]%
        {Riedl2012TheImaging.}
\bibfield{author}{\bibinfo{person}{René Riedl} {and} \bibinfo{person}{Andrija
  Javor}.} \bibinfo{year}{2012}\natexlab{}.
\newblock \showarticletitle{{The biology of trust: Integrating evidence from
  genetics, endocrinology, and functional brain imaging.}}
\newblock \bibinfo{journal}{\emph{Journal of Neuroscience, Psychology, and
  Economics}} \bibinfo{volume}{5}, \bibinfo{number}{2} (\bibinfo{date}{5}
  \bibinfo{year}{2012}), \bibinfo{pages}{63--91}.
\newblock
\showISSN{2151-318X}
\urldef\tempurl%
\url{https://doi.org/10.1037/a0026318}
\showDOI{\tempurl}


\bibitem[Rinck et~al\mbox{.}(2010)]%
        {Rinck2010AttentionalSpiders}
\bibfield{author}{\bibinfo{person}{Mike Rinck}, \bibinfo{person}{Linda
  Kwakkenbos}, \bibinfo{person}{Ron Dotsch}, \bibinfo{person}{Daniël~H.J.
  Wigboldus}, {and} \bibinfo{person}{Eni~S. Becker}.}
  \bibinfo{year}{2010}\natexlab{}.
\newblock \showarticletitle{{Attentional and behavioural responses of spider
  fearfuls to virtual spiders}}.
\newblock \bibinfo{journal}{\emph{Cognition and Emotion}}
  (\bibinfo{year}{2010}).
\newblock
\showISSN{02699931}
\urldef\tempurl%
\url{https://doi.org/10.1080/02699930903135945}
\showDOI{\tempurl}


\bibitem[Rosenberg(1965)]%
        {Rosenberg1965RosenbergScale}
\bibfield{author}{\bibinfo{person}{Morris Rosenberg}.}
  \bibinfo{year}{1965}\natexlab{}.
\newblock \showarticletitle{{Rosenberg self-esteem scale}}.
\newblock \bibinfo{journal}{\emph{Journal of Religion and Health}}
  (\bibinfo{year}{1965}).
\newblock


\bibitem[Rousseau et~al\mbox{.}(1998)]%
        {Rousseau1998NotTrust}
\bibfield{author}{\bibinfo{person}{Denise~M. Rousseau}, \bibinfo{person}{Sim~B.
  Sitkin}, \bibinfo{person}{Ronald~S. Burt}, {and} \bibinfo{person}{Colin
  Camerer}.} \bibinfo{year}{1998}\natexlab{}.
\newblock \showarticletitle{{Not So Different After All: A Cross-Discipline
  View Of Trust}}.
\newblock \bibinfo{journal}{\emph{Academy of Management Review}}
  \bibinfo{volume}{23}, \bibinfo{number}{3} (\bibinfo{date}{7}
  \bibinfo{year}{1998}), \bibinfo{pages}{393--404}.
\newblock
\showISSN{0363-7425}
\urldef\tempurl%
\url{https://doi.org/10.5465/amr.1998.926617}
\showDOI{\tempurl}


\bibitem[Ryan(2020)]%
        {Ryan2020InReliability}
\bibfield{author}{\bibinfo{person}{Mark Ryan}.}
  \bibinfo{year}{2020}\natexlab{}.
\newblock \showarticletitle{{In AI We Trust: Ethics, Artificial Intelligence,
  and Reliability}}.
\newblock \bibinfo{journal}{\emph{Science and Engineering Ethics}}
  \bibinfo{volume}{26}, \bibinfo{number}{5} (\bibinfo{date}{10}
  \bibinfo{year}{2020}), \bibinfo{pages}{2749--2767}.
\newblock
\showISSN{1353-3452}
\urldef\tempurl%
\url{https://doi.org/10.1007/s11948-020-00228-y}
\showDOI{\tempurl}


\bibitem[Salkovskis(1991)]%
        {Salkovskis1991TheAccount}
\bibfield{author}{\bibinfo{person}{Paul~M. Salkovskis}.}
  \bibinfo{year}{1991}\natexlab{}.
\newblock \showarticletitle{{The Importance of Behaviour in the Maintenance of
  Anxiety and Panic: A Cognitive Account}}.
\newblock \bibinfo{journal}{\emph{Behavioural Psychotherapy}}
  (\bibinfo{year}{1991}).
\newblock
\showISSN{01413473}
\urldef\tempurl%
\url{https://doi.org/10.1017/S0141347300011472}
\showDOI{\tempurl}


\bibitem[Schranz et~al\mbox{.}(2018)]%
        {Schranz2018MediaUse}
\bibfield{author}{\bibinfo{person}{Mario Schranz}, \bibinfo{person}{Jörg
  Schneider}, {and} \bibinfo{person}{Mark Eisenegger}.}
  \bibinfo{year}{2018}\natexlab{}.
\newblock \showarticletitle{{Media Trust and Media Use}}.
\newblock In \bibinfo{booktitle}{\emph{Trust in Media and Journalism}}.
  \bibinfo{publisher}{Springer Fachmedien Wiesbaden},
  \bibinfo{address}{Wiesbaden}, \bibinfo{pages}{73--91}.
\newblock
\urldef\tempurl%
\url{https://doi.org/10.1007/978-3-658-20765-6{\_}5}
\showDOI{\tempurl}


\bibitem[See et~al\mbox{.}(2011)]%
        {See2011TheAccuracy}
\bibfield{author}{\bibinfo{person}{Kelly~E. See}, \bibinfo{person}{Elizabeth~W.
  Morrison}, \bibinfo{person}{Naomi~B. Rothman}, {and} \bibinfo{person}{Jack~B.
  Soll}.} \bibinfo{year}{2011}\natexlab{}.
\newblock \showarticletitle{{The detrimental effects of power on confidence,
  advice taking, and accuracy}}.
\newblock \bibinfo{journal}{\emph{Organizational Behavior and Human Decision
  Processes}} \bibinfo{volume}{116}, \bibinfo{number}{2} (\bibinfo{date}{11}
  \bibinfo{year}{2011}), \bibinfo{pages}{272--285}.
\newblock
\showISSN{07495978}
\urldef\tempurl%
\url{https://doi.org/10.1016/j.obhdp.2011.07.006}
\showDOI{\tempurl}


\bibitem[Solberg et~al\mbox{.}(2022)]%
        {Solberg2022AAids}
\bibfield{author}{\bibinfo{person}{Elizabeth Solberg},
  \bibinfo{person}{Magnhild Kaarstad}, \bibinfo{person}{Maren~H.Rø Eitrheim},
  \bibinfo{person}{Rossella Bisio}, \bibinfo{person}{Kine Reeg{\aa}rd}, {and}
  \bibinfo{person}{Marten Bloch}.} \bibinfo{year}{2022}\natexlab{}.
\newblock \showarticletitle{{A Conceptual Model of Trust, Perceived Risk, and
  Reliance on AI Decision Aids}}.
\newblock \bibinfo{journal}{\emph{Group and Organization Management}}
  \bibinfo{volume}{47}, \bibinfo{number}{2} (\bibinfo{date}{4}
  \bibinfo{year}{2022}), \bibinfo{pages}{187--222}.
\newblock
\showISSN{15523993}
\urldef\tempurl%
\url{https://doi.org/10.1177/10596011221081238}
\showDOI{\tempurl}


\bibitem[S{\"{o}}llner et~al\mbox{.}(2016)]%
        {Sollner2016WhyUsers}
\bibfield{author}{\bibinfo{person}{Matthias S{\"{o}}llner},
  \bibinfo{person}{Axel Hoffmann}, {and} \bibinfo{person}{Jan~Marco
  Leimeister}.} \bibinfo{year}{2016}\natexlab{}.
\newblock \showarticletitle{{Why different trust relationships matter for
  information systems users}}.
\newblock \bibinfo{journal}{\emph{European Journal of Information Systems}}
  \bibinfo{volume}{25}, \bibinfo{number}{3} (\bibinfo{date}{5}
  \bibinfo{year}{2016}), \bibinfo{pages}{274--287}.
\newblock
\showISSN{0960-085X}
\urldef\tempurl%
\url{https://doi.org/10.1057/ejis.2015.17}
\showDOI{\tempurl}


\bibitem[Spence and Rapee(2016)]%
        {Spence2016}
\bibfield{author}{\bibinfo{person}{Susan~H. Spence} {and}
  \bibinfo{person}{Ronald~M. Rapee}.} \bibinfo{year}{2016}\natexlab{}.
\newblock \bibinfo{title}{{The etiology of social anxiety disorder: An
  evidence-based model}}.
\newblock
\newblock
\showISSN{1873622X}
\urldef\tempurl%
\url{https://doi.org/10.1016/j.brat.2016.06.007}
\showDOI{\tempurl}


\bibitem[Strimbu and Tavel(2010)]%
        {Strimbu2010}
\bibfield{author}{\bibinfo{person}{Kyle Strimbu} {and} \bibinfo{person}{Jorge~A
  Tavel}.} \bibinfo{year}{2010}\natexlab{}.
\newblock \showarticletitle{{What are biomarkers?}}
\newblock \bibinfo{journal}{\emph{Current opinion in HIV and AIDS}}
  \bibinfo{volume}{5}, \bibinfo{number}{6} (\bibinfo{date}{11}
  \bibinfo{year}{2010}), \bibinfo{pages}{463--466}.
\newblock
\showISSN{1746-6318}
\urldef\tempurl%
\url{https://doi.org/10.1097/COH.0b013e32833ed177}
\showDOI{\tempurl}


\bibitem[Sutcliffe et~al\mbox{.}(2015)]%
        {Sutcliffe2015ModellingRelationships}
\bibfield{author}{\bibinfo{person}{Alistair~G. Sutcliffe}, \bibinfo{person}{Di
  Wang}, {and} \bibinfo{person}{Robin I.~M. Dunbar}.}
  \bibinfo{year}{2015}\natexlab{}.
\newblock \showarticletitle{{Modelling the Role of Trust in Social
  Relationships}}.
\newblock \bibinfo{journal}{\emph{ACM Transactions on Internet Technology}}
  \bibinfo{volume}{15}, \bibinfo{number}{4} (\bibinfo{date}{12}
  \bibinfo{year}{2015}), \bibinfo{pages}{1--24}.
\newblock
\showISSN{1533-5399}
\urldef\tempurl%
\url{https://doi.org/10.1145/2815620}
\showDOI{\tempurl}


\bibitem[Sutherland et~al\mbox{.}(2015)]%
        {Sutherland2015TheAutomation}
\bibfield{author}{\bibinfo{person}{Steven~C. Sutherland},
  \bibinfo{person}{Casper Harteveld}, {and} \bibinfo{person}{Michael~E.
  Young}.} \bibinfo{year}{2015}\natexlab{}.
\newblock \showarticletitle{{The Role of Environmental Predictability and Costs
  in Relying on Automation}}. In \bibinfo{booktitle}{\emph{Proceedings of the
  33rd Annual ACM Conference on Human Factors in Computing Systems}}.
  \bibinfo{publisher}{ACM}, \bibinfo{address}{New York, NY, USA},
  \bibinfo{pages}{2535--2544}.
\newblock
\showISBNx{9781450331456}
\urldef\tempurl%
\url{https://doi.org/10.1145/2702123.2702609}
\showDOI{\tempurl}


\bibitem[Tolmeijer et~al\mbox{.}(2022)]%
        {Tolmeijer2022CapableMaking}
\bibfield{author}{\bibinfo{person}{Suzanne Tolmeijer}, \bibinfo{person}{Markus
  Christen}, \bibinfo{person}{Serhiy Kandul}, \bibinfo{person}{Markus Kneer},
  {and} \bibinfo{person}{Abraham Bernstein}.} \bibinfo{year}{2022}\natexlab{}.
\newblock \showarticletitle{{Capable but Amoral? Comparing AI and Human Expert
  Collaboration in Ethical Decision Making}}. In \bibinfo{booktitle}{\emph{CHI
  Conference on Human Factors in Computing Systems}}. \bibinfo{publisher}{ACM},
  \bibinfo{address}{New York, NY, USA}, \bibinfo{pages}{1--17}.
\newblock
\showISBNx{9781450391573}
\urldef\tempurl%
\url{https://doi.org/10.1145/3491102.3517732}
\showDOI{\tempurl}


\bibitem[Ueno et~al\mbox{.}(2022)]%
        {Ueno2022TrustMethods}
\bibfield{author}{\bibinfo{person}{Takane Ueno}, \bibinfo{person}{Yuto Sawa},
  \bibinfo{person}{Yeongdae Kim}, \bibinfo{person}{Jacqueline Urakami},
  \bibinfo{person}{Hiroki Oura}, {and} \bibinfo{person}{Katie Seaborn}.}
  \bibinfo{year}{2022}\natexlab{}.
\newblock \showarticletitle{{Trust in Human-AI Interaction: Scoping Out Models,
  Measures, and Methods}}. In \bibinfo{booktitle}{\emph{CHI Conference on Human
  Factors in Computing Systems Extended Abstracts}}. \bibinfo{publisher}{ACM},
  \bibinfo{address}{New York, NY, USA}, \bibinfo{pages}{1--7}.
\newblock
\showISBNx{9781450391566}
\urldef\tempurl%
\url{https://doi.org/10.1145/3491101.3519772}
\showDOI{\tempurl}


\bibitem[{Unity Technologies}(2020)]%
        {UnityTechnologies2020}
\bibfield{author}{\bibinfo{person}{{Unity Technologies}}.}
  \bibinfo{year}{2020}\natexlab{}.
\newblock \bibinfo{title}{{Unity Engine}}.
\newblock
\newblock
\urldef\tempurl%
\url{https://unity.com/}
\showURL{%
\tempurl}


\bibitem[van~der Waa et~al\mbox{.}(2018)]%
        {vanderWaa2018ContrastiveConsequences}
\bibfield{author}{\bibinfo{person}{Jasper van~der Waa},
  \bibinfo{person}{Jurriaan van Diggelen}, \bibinfo{person}{Karel van~den
  Bosch}, {and} \bibinfo{person}{Mark Neerincx}.}
  \bibinfo{year}{2018}\natexlab{}.
\newblock \showarticletitle{{Contrastive Explanations for Reinforcement
  Learning in terms of Expected Consequences}}.
\newblock  (\bibinfo{date}{7} \bibinfo{year}{2018}).
\newblock


\bibitem[van Maanen et~al\mbox{.}(2011)]%
        {vanMaanen2011EffectsAutonomy}
\bibfield{author}{\bibinfo{person}{Peter-Paul van Maanen},
  \bibinfo{person}{Francien Wisse}, \bibinfo{person}{Jurriaan van Diggelen},
  {and} \bibinfo{person}{Robbert-Jan Beun}.} \bibinfo{year}{2011}\natexlab{}.
\newblock \showarticletitle{{Effects of Reliance Support on Team Performance by
  Advising and Adaptive Autonomy}}. In \bibinfo{booktitle}{\emph{2011
  IEEE/WIC/ACM International Conferences on Web Intelligence and Intelligent
  Agent Technology}}. \bibinfo{publisher}{IEEE}, \bibinfo{pages}{280--287}.
\newblock
\showISBNx{978-1-4577-1373-6}
\urldef\tempurl%
\url{https://doi.org/10.1109/WI-IAT.2011.117}
\showDOI{\tempurl}


\bibitem[Vereschak et~al\mbox{.}(2021)]%
        {Vereschak2021HowMethodologies}
\bibfield{author}{\bibinfo{person}{Oleksandra Vereschak},
  \bibinfo{person}{Gilles Bailly}, {and} \bibinfo{person}{Baptiste Caramiaux}.}
  \bibinfo{year}{2021}\natexlab{}.
\newblock \showarticletitle{{How to Evaluate Trust in AI-Assisted Decision
  Making? A Survey of Empirical Methodologies}}.
\newblock \bibinfo{journal}{\emph{Proceedings of the ACM on Human-Computer
  Interaction}} \bibinfo{volume}{5}, \bibinfo{number}{CSCW2}
  (\bibinfo{date}{10} \bibinfo{year}{2021}).
\newblock
\showISSN{25730142}
\urldef\tempurl%
\url{https://doi.org/10.1145/3476068}
\showDOI{\tempurl}


\bibitem[Wrightsman(1991)]%
        {Wrightsman_1991}
\bibfield{author}{\bibinfo{person}{Lawrence~S Wrightsman}.}
  \bibinfo{year}{1991}\natexlab{}.
\newblock \showarticletitle{{Interpersonal trust and attitudes toward human
  nature.}}
\newblock  (\bibinfo{year}{1991}).
\newblock


\bibitem[Wu et~al\mbox{.}(2022)]%
        {Wu2022SafetyNavigation}
\bibfield{author}{\bibinfo{person}{Jianjun Wu}, \bibinfo{person}{James
  Thorne-Large}, {and} \bibinfo{person}{Pengfei Zhang}.}
  \bibinfo{year}{2022}\natexlab{}.
\newblock \showarticletitle{{Safety first: The risk of over-reliance on
  technology in navigation}}.
\newblock \bibinfo{journal}{\emph{Journal of Transportation Safety {\&}
  Security}} \bibinfo{volume}{14}, \bibinfo{number}{7} (\bibinfo{date}{7}
  \bibinfo{year}{2022}), \bibinfo{pages}{1220--1246}.
\newblock
\showISSN{1943-9962}
\urldef\tempurl%
\url{https://doi.org/10.1080/19439962.2021.1909681}
\showDOI{\tempurl}


\bibitem[Xiao and Seagull(1999)]%
        {Xiao1999AnTasks}
\bibfield{author}{\bibinfo{person}{Yan Xiao} {and} \bibinfo{person}{F.~Jacob
  Seagull}.} \bibinfo{year}{1999}\natexlab{}.
\newblock \showarticletitle{{An Analysis of Problems with Auditory Alarms:
  Defining the Roles of Alarms in Process Monitoring Tasks}}.
\newblock \bibinfo{journal}{\emph{Proceedings of the Human Factors and
  Ergonomics Society Annual Meeting}} \bibinfo{volume}{43}, \bibinfo{number}{3}
  (\bibinfo{date}{9} \bibinfo{year}{1999}), \bibinfo{pages}{256--260}.
\newblock
\showISSN{2169-5067}
\urldef\tempurl%
\url{https://doi.org/10.1177/154193129904300327}
\showDOI{\tempurl}


\bibitem[Yamauchi(2013)]%
        {Yamauchi2013MouseForest}
\bibfield{author}{\bibinfo{person}{Takashi Yamauchi}.}
  \bibinfo{year}{2013}\natexlab{}.
\newblock \showarticletitle{{Mouse Trajectories and State Anxiety: Feature
  Selection with Random Forest}}. In \bibinfo{booktitle}{\emph{2013 Humaine
  Association Conference on Affective Computing and Intelligent Interaction}}.
  \bibinfo{publisher}{IEEE}, \bibinfo{pages}{399--404}.
\newblock
\showISBNx{978-0-7695-5048-0}
\urldef\tempurl%
\url{https://doi.org/10.1109/ACII.2013.72}
\showDOI{\tempurl}


\bibitem[Yi et~al\mbox{.}(2020)]%
        {Yi2020IdentificationEmotions}
\bibfield{author}{\bibinfo{person}{Qian Yi}, \bibinfo{person}{Shiquan Xiong},
  \bibinfo{person}{Biao Wang}, {and} \bibinfo{person}{Shuping Yi}.}
  \bibinfo{year}{2020}\natexlab{}.
\newblock \showarticletitle{{Identification of trusted interactive behavior
  based on mouse behavior considering web User's emotions}}.
\newblock \bibinfo{journal}{\emph{International Journal of Industrial
  Ergonomics}}  \bibinfo{volume}{76} (\bibinfo{date}{3} \bibinfo{year}{2020}),
  \bibinfo{pages}{102903}.
\newblock
\showISSN{01698141}
\urldef\tempurl%
\url{https://doi.org/10.1016/j.ergon.2019.102903}
\showDOI{\tempurl}


\bibitem[Yin et~al\mbox{.}(2019)]%
        {Yin2019UnderstandingModels}
\bibfield{author}{\bibinfo{person}{Ming Yin}, \bibinfo{person}{Jennifer
  Wortman~Vaughan}, {and} \bibinfo{person}{Hanna Wallach}.}
  \bibinfo{year}{2019}\natexlab{}.
\newblock \showarticletitle{{Understanding the Effect of Accuracy on Trust in
  Machine Learning Models}}. In \bibinfo{booktitle}{\emph{Proceedings of the
  2019 CHI Conference on Human Factors in Computing Systems}}.
  \bibinfo{publisher}{ACM}, \bibinfo{address}{New York, NY, USA},
  \bibinfo{pages}{1--12}.
\newblock
\showISBNx{9781450359702}
\urldef\tempurl%
\url{https://doi.org/10.1145/3290605.3300509}
\showDOI{\tempurl}


\bibitem[Yu et~al\mbox{.}(2019)]%
        {Yu2019DoDecision}
\bibfield{author}{\bibinfo{person}{Kun Yu}, \bibinfo{person}{Shlomo Berkovsky},
  \bibinfo{person}{Ronnie Taib}, \bibinfo{person}{Jianlong Zhou}, {and}
  \bibinfo{person}{Fang Chen}.} \bibinfo{year}{2019}\natexlab{}.
\newblock \showarticletitle{{Do I trust my machine teammate?: an investigation
  from perception to decision}}. In \bibinfo{booktitle}{\emph{Proceedings of
  the 24th International Conference on Intelligent User Interfaces}}.
  \bibinfo{publisher}{ACM}, \bibinfo{address}{New York, NY, USA},
  \bibinfo{pages}{460--468}.
\newblock
\showISBNx{9781450362726}
\urldef\tempurl%
\url{https://doi.org/10.1145/3301275.3302277}
\showDOI{\tempurl}


\bibitem[Yu et~al\mbox{.}(2018)]%
        {Yu2018ArtificialHealthcare}
\bibfield{author}{\bibinfo{person}{Kun-Hsing Yu}, \bibinfo{person}{Andrew~L.
  Beam}, {and} \bibinfo{person}{Isaac~S. Kohane}.}
  \bibinfo{year}{2018}\natexlab{}.
\newblock \showarticletitle{{Artificial intelligence in healthcare}}.
\newblock \bibinfo{journal}{\emph{Nature Biomedical Engineering}}
  \bibinfo{volume}{2}, \bibinfo{number}{10} (\bibinfo{date}{10}
  \bibinfo{year}{2018}), \bibinfo{pages}{719--731}.
\newblock
\showISSN{2157-846X}
\urldef\tempurl%
\url{https://doi.org/10.1038/s41551-018-0305-z}
\showDOI{\tempurl}


\bibitem[Yuksel et~al\mbox{.}(2017)]%
        {Yuksel2017BrainsInteractions}
\bibfield{author}{\bibinfo{person}{Beste~F. Yuksel}, \bibinfo{person}{Penny
  Collisson}, {and} \bibinfo{person}{Mary Czerwinski}.}
  \bibinfo{year}{2017}\natexlab{}.
\newblock \showarticletitle{{Brains or Beauty: How to Engender Trust in
  User-Agent Interactions}}.
\newblock \bibinfo{journal}{\emph{ACM Transactions on Internet Technology}}
  \bibinfo{volume}{17}, \bibinfo{number}{1} (\bibinfo{date}{3}
  \bibinfo{year}{2017}), \bibinfo{pages}{1--20}.
\newblock
\showISSN{1533-5399}
\urldef\tempurl%
\url{https://doi.org/10.1145/2998572}
\showDOI{\tempurl}


\bibitem[Zaheer et~al\mbox{.}(1998)]%
        {Zaheer1998DoesPerformance}
\bibfield{author}{\bibinfo{person}{Akbar Zaheer}, \bibinfo{person}{Bill
  McEvily}, {and} \bibinfo{person}{Vincenzo Perrone}.}
  \bibinfo{year}{1998}\natexlab{}.
\newblock \showarticletitle{{Does Trust Matter? Exploring the Effects of
  Interorganizational and Interpersonal Trust on Performance}}.
\newblock \bibinfo{journal}{\emph{Organization Science}} \bibinfo{volume}{9},
  \bibinfo{number}{2} (\bibinfo{date}{4} \bibinfo{year}{1998}),
  \bibinfo{pages}{141--159}.
\newblock
\showISSN{1047-7039}
\urldef\tempurl%
\url{https://doi.org/10.1287/orsc.9.2.141}
\showDOI{\tempurl}


\bibitem[Zhang et~al\mbox{.}(2020)]%
        {Zhang2020EffectMaking}
\bibfield{author}{\bibinfo{person}{Yunfeng Zhang}, \bibinfo{person}{Q.~Vera
  Liao}, {and} \bibinfo{person}{Rachel K.~E. Bellamy}.}
  \bibinfo{year}{2020}\natexlab{}.
\newblock \showarticletitle{{Effect of confidence and explanation on accuracy
  and trust calibration in AI-assisted decision making}}. In
  \bibinfo{booktitle}{\emph{Proceedings of the 2020 Conference on Fairness,
  Accountability, and Transparency}}. \bibinfo{publisher}{ACM},
  \bibinfo{address}{New York, NY, USA}, \bibinfo{pages}{295--305}.
\newblock
\showISBNx{9781450369367}
\urldef\tempurl%
\url{https://doi.org/10.1145/3351095.3372852}
\showDOI{\tempurl}


\bibitem[Zhu et~al\mbox{.}(2019)]%
        {Zhu2019RivalsHiding}
\bibfield{author}{\bibinfo{person}{Yue Zhu}, \bibinfo{person}{Tingting Chen},
  \bibinfo{person}{Mo Wang}, \bibinfo{person}{Yanghua Jin}, {and}
  \bibinfo{person}{Yongyue Wang}.} \bibinfo{year}{2019}\natexlab{}.
\newblock \showarticletitle{{Rivals or allies: How performance‐prove goal
  orientation influences knowledge hiding}}.
\newblock \bibinfo{journal}{\emph{Journal of Organizational Behavior}}
  \bibinfo{volume}{40}, \bibinfo{number}{7} (\bibinfo{date}{9}
  \bibinfo{year}{2019}), \bibinfo{pages}{849--868}.
\newblock
\showISSN{0894-3796}
\urldef\tempurl%
\url{https://doi.org/10.1002/job.2372}
\showDOI{\tempurl}


\bibitem[Zushi et~al\mbox{.}(2015)]%
        {Zushi2015AnalysisProblems}
\bibfield{author}{\bibinfo{person}{Mitsumasa Zushi}, \bibinfo{person}{Yoshinori
  Miyazaki}, {and} \bibinfo{person}{Ken Norizuki}.}
  \bibinfo{year}{2015}\natexlab{}.
\newblock \showarticletitle{{Analysis of learners' study logs: mouse
  trajectories to identify the occurrence of hesitation in solving
  word-reordering problems}}. In \bibinfo{booktitle}{\emph{Proceedings of the
  Fifth International Conference on Learning Analytics And Knowledge}}.
  \bibinfo{publisher}{ACM}, \bibinfo{address}{New York, NY, USA},
  \bibinfo{pages}{428--429}.
\newblock
\showISBNx{9781450334174}
\urldef\tempurl%
\url{https://doi.org/10.1145/2723576.2723645}
\showDOI{\tempurl}


\end{thebibliography}

\end{document}